\documentclass{emulateapj}
\usepackage{graphicx}
\usepackage{bm}
\usepackage{amssymb}
\usepackage{gensymb}
\usepackage{amsmath}
\usepackage{commath}
\usepackage{cancel}

\usepackage{epstopdf}
\usepackage{natbib}
\usepackage{epsfig}

\usepackage{ulem}
\usepackage{lineno}

\newcommand{\avg}[1]{\ensuremath{\langle #1 \rangle}}
\newcommand{\bma}{\begin{math}}
\newcommand{\ema}{\end{math}}
\newcommand{\beq}{\begin{equation}}
\newcommand{\eeq}{\end{equation}}
\newcommand{\beqa}{\begin{eqnarray}}
\newcommand{\eeqa}{\end{eqnarray}}
\newcommand{\bc}{\begin{center}}
\newcommand{\ec}{\end{center}} 
\newcommand{\bit}{\begin{itemize}}
\newcommand{\eit}{\end{itemize}}

\font\BFd=cmmib10
\font\BFt=cmmib10
\font\BFs=cmmib10 scaled 700
\font\BFss=cmmib10 scaled 500

\def\bbox#1{%
\relax\ifmmode
\mathchoice
{{\hbox{\BFd #1}}}
{{\hbox{\BFt #1}}}
{{\hbox{\BFs #1}}}
{{\hbox{\BFss #1}}}
\else \mbox{#1} \fi }

\def\k{{\bbox{k}}}

\usepackage{etoolbox}
\usepackage{xcolor}

\newtoggle{commentsoff}
\iftoggle{commentsoff}{
	\newcommand{\dhj}[1]{}
	\newcommand{\al}[1]{}
	\newcommand{\ab}[1]{}
	\newcommand{\sp}[1]{}
	\newcommand{\rc}[1]{}
}{
\newcommand{\dhj}[1]{{\color{teal}~\textsf{[{\bf Dana}: #1]}}}
\newcommand{\al}[1]{{\color{red}~\textsf{[{\bf Adam}: #1]}}}
\newcommand{\ab}[1]{{\color{blue}~\textsf{[{\bf Gus}: #1]}}}

\newcommand{\rc}[1]{{\color{violet}~\textsf{[{\bf Richard}: #1]}}}
}

\begin{document}

\title{Fuzzy Dark Matter and the 21 cm Power Spectrum}
\author{Dana Jones$^{1}$, Skyler Palatnick$^{1}$, Richard Chen$^{1}$, Angus Beane$^{2}$, Adam Lidz$^{1}$}
\altaffiltext{1} {Department of Physics \& Astronomy, University of Pennsylvania, 209 South 33rd Street, Philadelphia, PA 19104, USA}
\altaffiltext{2} {Center for Astrophysics {\normalfont |} Harvard \& Smithsonian, 60 Garden Street, Cambridge, MA 02138, USA}
\email{dhjones@sas.upenn.edu,skylerp@sas.upenn.edu}

\begin{abstract}
We model the 21 cm power spectrum across the Cosmic Dawn and the Epoch of Reionization (EoR) in fuzzy dark matter (FDM) cosmologies. The suppression of small mass halos in FDM models leads to a delay in the onset redshift of these epochs relative to cold dark matter (CDM) scenarios. This strongly impacts the 21 cm power spectrum and its redshift evolution. The 21 cm power spectrum at a given stage -- i.e., compared at fixed average brightness temperature but varying redshift -- of the EoR/Cosmic Dawn process is also modified: in general, the amplitude of 21 cm fluctuations is boosted by the enhanced bias factor of galaxy hosting halos in FDM. 
We forecast the prospects for discriminating between CDM and FDM with upcoming power spectrum measurements from HERA, accounting for degeneracies between astrophysical parameters and dark matter properties. If FDM constitutes the entirety of the dark matter and the FDM particle mass is $10^{-21}$ eV, HERA can determine the mass to within 20\% at
$2-\sigma$ confidence. 
\end{abstract}



\section{Introduction} \label{sec:intro}

In spite of decades of effort, the particle properties of dark matter remain mysterious.  One well-motivated possibility is that the dark matter consists of elementary particles with weak-scale interaction cross sections and particle masses (i.e., masses of order 100 GeV or thereabouts); these particles were produced thermally in the early universe and are non-relativistic during structure formation, behaving as cold dark matter (CDM). However, direct detection experiments, collider searches, and indirect methods have yet to make convincing detections and have placed increasingly stringent bounds on these weakly-interacting massive particle (WIMP) candidates (see e.g. the review by \citealt{Arcadi:2017kky}). Although regions of parameter space remain unconstrained, and hence it is still feasible that WIMPs make up the entirety of the dark matter, the recent limits have further motivated the study of alternative possibilities. 

Among these, an intriguing case is that of fuzzy dark matter \citep[FDM; ][]{Hu:2000ke}. In FDM the dark matter consists of extremely light scalar particles with masses of order $m_{\mathrm{FDM}} \sim 10^{-22}$ eV. This possibility is well-motivated by the ubiquitous presence of ultralight scalar fields in theories beyond the standard model of particle physics, while  the present day dark matter abundance may be naturally obtained for this general mass range \citep{Hui:2016ltb}. Furthermore, FDM has distinctive astrophysical signatures that may allow one to confirm or rule-out its presence. In particular, the small particle mass in FDM gives rise to macroscopic DeBroglie wavelengths, which can be $\sim$ kpc in scale depending on the particle velocity, and this leads to a host of interesting astrophysical consequences. In general, FDM preserves the well-established success of CDM on large-scales, while providing different predictions on small-scales \citep{Hui:2016ltb}.

One consequence of the macroscopic DeBroglie wavelengths in FDM is that the power spectrum of initial density fluctuations is truncated on small scales \citep{Hu:2000ke}. This strongly suppresses the abundance of small mass dark matter halos relative to the case of CDM. This should, in turn, lead to delays in the earliest phases of galaxy formation; in CDM small mass halos collapse first and galaxies form as gas subsequently falls into the dark matter potential wells, cools, and fragments to form stars \citep{Gunn78,White78}. In FDM this process is delayed until halos above the suppression mass start to collapse. A promising way of testing FDM is therefore to study the Epoch of Reionization (EoR) and Cosmic Dawn eras when the first galaxies form, emit ultraviolet light, and gradually photoionize and heat the surrounding intergalactic medium (IGM) \citep{Hu:2000ke,Bozek:2014uqa,Lidz:2018fqo}. 

One of the most exciting ways to probe the EoR and Cosmic Dawn eras is via the redshifted 21 cm line \citep{Furlanetto:2006jb,Pritchard2012}. First, as early sources of radiation turn on, a background of Ly-$\alpha$ photons builds up and couples the spin temperature of the 21 cm line to the gas temperature. At this time the gas temperature is expected to be less than the cosmic microwave background (CMB) temperature, and the 21 cm signal should be observable in absorption relative to the CMB. Subsequently, early sources of X-ray emission raise the gas temperature above the CMB temperature. These processes are expected to occur before most of the IGM is reionized and these earliest phases -- before the bulk of reionization -- are hence referred to as the ``Cosmic Dawn''.\footnote{In some parts of the literature, ``Cosmic Dawn'' and reionization are treated as synonymous. Here we prefer a 21 cm-centric definition in which Cosmic Dawn refers to the Ly-$\alpha$ coupling and X-ray heating phases, while we use the EoR to denote subsequent stages of reionization after X-ray heating is complete.} Finally, ionized regions around the first luminous sources gradually grow, merge, and eventually fill essentially the entire volume of the universe during the EoR. The overall timing of this process, and its statistical properties, should reveal the nature of the first luminous objects and also provide a powerful test of dark matter properties, with FDM having a potentially dramatic impact. 

In fact, the EDGES collaboration recently reported evidence of a feature in the sky-averaged radio spectrum which they interpreted as a signature of 21 cm absorption at $z \sim 15-20$ \citep{Bowman:2018yin}. Taken at face value, this detection of the global average 21 cm signal implies an early start to structure formation and that a Ly-$\alpha$ background was already established by $z \sim 15-20$. This, in turn, leads to tight limits on the possibility that FDM makes up the entirety of the dark matter \citep{Schneider:2018xba,Lidz:2018fqo,Nebrin:2018vqt}. However, the EDGES signal has puzzling features (see e.g. \citealt{Mirocha:2018cih,Lidz:2018fqo,Schauer:2019ihk,Fialkov:2019vnb,Reis:2020arr}). Moreover, the global average 21 cm signal is a challenging measurement and a number of works have pointed out concerns with the EDGES analysis (e.g. \citealt{Hills:2018vyr,Singh19,Sims:2019kro}). 

In addition to the global average 21 cm signal, it may be possible to measure spatial fluctuations in the 21 cm signal across the sky and as a function of frequency. Indeed, ongoing and upcoming projects aim to measure the power spectrum of 21 cm fluctuations, eventually spanning the entire redshift range of the EoR, the Cosmic Dawn, and ultimately the preceding dark ages \citep{Furlanetto:2006jb}. These measurements have a different set of systematic concerns than the global 21 cm experiments, and offer a potentially richer data set to exploit. 
Especially exciting in this regard is the HERA survey, which is underway, and is forecasted to measure the 21 cm power spectrum at high statistical significance across a broad range of redshifts \citep{DeBoer:2016tnn}. In the future, the SKA \citep{Dewdney:2009} should provide even more precise measurements.

The goal of this paper is to model the 21 cm fluctuations during the EoR and Cosmic Dawn in FDM, to characterize the differences with CDM models, and to forecast the prospects for detecting or constraining FDM with HERA, while exploring degeneracies with some of the uncertain astrophysical parameters involved. 
We use the publicly available 21cmFAST code \citep{Mesinger11} to model reionization and Cosmic Dawn, and a Fisher matrix formalism to forecast the constraining power of HERA. 
In \S \ref{sec:method} we describe our models. \S \ref{sec:qualitative_results} provides a qualitative description of the impact of FDM on the redshifted 21 cm signal, while \S \ref{sec:HERA} quantifies the sensitivity of HERA and its prospects for discriminating  between CDM and FDM. These results are sharpened in \S \ref{sec:forecasts}, where we present full Fisher matrix forecasts. Finally, we conclude and discuss possible future directions in \S \ref{sec:conclusions}. 

In considering the 21 cm power spectrum in FDM, this study has some overlap with earlier work by \cite{Sitwell14,Munoz20,Nebrin:2018vqt} who investigated the 21 cm power spectrum in FDM and/or the related case of warm dark matter (WDM) models. Although WDM and FDM are physically very different models for the dark matter, they each lead to a suppression in the power spectrum of initial density fluctuations and delay reionization/Cosmic Dawn. We focus on the FDM case here, but translate our results into WDM constraints in the Conclusion.
Our independent analysis includes full Fisher forecasts for HERA and furthers the discussion in \cite{Sitwell14} and \cite{Nebrin:2018vqt} which did not include such forecasts. The more recent work by \citet{Munoz20} does include Fisher forecasts for HERA measurements, and is broadly consistent with our work, although these authors adopt a slightly different approach and emphasis. 
Throughout we assume the following cosmological parameters, based on Planck 2015 constraints and consistent with Planck 2018 results \citep{Aghanim:2018eyx}: $(\Omega_m, \Omega_b h^2, \Omega_\Lambda, h, \sigma_8, n_s) = (0.308, 0.02226, 0.691, 0.678, 0.815, 0.968)$, where these have their usual meanings and $\sigma_8$ and $n_s$ describe linear power spectrum in the CDM case.

\section{Reionization and Cosmic Dawn Models}\label{sec:method}

Here we briefly describe the 21 cm signal and the simulations used to model Cosmic Dawn/the EoR in CDM and FDM. 
The 21 cm brightness temperature contrast of a neutral hydrogen cloud, at co-moving spatial position $\bm{x}$, relative to the cosmic microwave background (CMB) is given by \citep{Madau:1996cs}:
\beq\label{eq:tbright}
T_{21}(\bm{x}) = T_0 x_{\rm{HI}}(\bm{x}) \left[\frac{T_s(\bm{x}) - T_\gamma}{T_s(\bm{x})}\right] \left[1 + \delta_\rho({\bm{x}})\right] \text{.}
\eeq
Here $T_0$ is a normalization constant, $T_0 = 28 {\rm mK} \left[(1+z)/10\right]^{1/2}$, $x_{\rm{HI}}$ is the neutral fraction of hydrogen, $T_s$ is the spin temperature of the 21 cm transition, $T_\gamma$ is the temperature of the radio background (which we assume throughout is dominated by the CMB),  and $1+\delta_\rho$ is the gas density in units of the cosmic mean. The gas density fluctuations are assumed to trace the overall matter density variations on the scales of interest. Note that we model spin temperature fluctuations in our analysis, as these produce strong spatial variations during the Cosmic Dawn era, and so $T_s$ is a function of spatial position in Eq~\ref{eq:tbright}. All of the quantities here generally evolve strongly with redshift, but the $z$ dependence is suppressed in the above equation for brevity.
For simplicity, we ignore the impact of peculiar velocities throughout this work (see e.g. \citealt{2012MNRAS.422..926M}).

The main observable of interest for our current study is the power spectrum of 21 cm brightness temperature fluctuations, $P_{21}(k)$. This power spectrum is defined by
\beq\label{eq:p21_definition}
\langle T_{21}(\bm{k}) T_{21}(\bm{k^\prime}) \rangle = (2 \pi)^3 \delta_D(\bm{k} + \bm{k^\prime}) P_{21}(k) \text{,}
\eeq
where $\delta_D$ denotes a Dirac delta function. We generally work with the related quantity, $\Delta^2_{21}(k) = k^3 P_{21}(k)/(2 \pi^2)$, which gives the variance of the 21 cm brightness temperature fluctuations per $\rm{ln}(k)$ with our Fourier convention. Throughout, we describe $\Delta^2_{21}(k)$ in units of $\rm{mK}^2$.

\subsection{21cmFAST simulations}

In order to model the Cosmic Dawn and reionization, we make use of the 21cmFAST code \citep{Mesinger11}, specifically version 1.3. 21cmFAST produces ``semi-numerical'' realizations of the reionization process \citep{Zahn:2006sg,Mesinger:2007pd}, based on an excursion-set formalism \citep{Bond91} for reionization \citep{Furlanetto:2004nh}. The code also includes an approximate treatment of Ly-$\alpha$ background photons, responsible for coupling the spin temperature of the 21 cm line to the gas temperature, and of X-ray heating. 

The simulations employed in our study span 300 Mpc co-moving on a side, and the density, ioniziation, and 21 cm fields are produced on a $256^3$ grid. Each 21cmFAST model is characterized by several ``astrophysical parameters'' describing the properties of the ionizing sources, and their production of UV and X-ray photons. First, $\zeta$ is a parameter describing the efficiency of ionizing photon production, which we set to $\zeta=20$. Second, $T_{\rm vir}$ is the minimum virial temperature of galaxy hosting dark matter halos. We adopt $T_{\rm vir} = 2 \times 10^4$ K for our fiducial model. Third, the maximum smoothing scale adopted in generating the ionization field is taken to be $R_{\rm max} = 50$ co-moving Mpc. We neglect any redshift dependence in these parameters and in those that follow. The model star formation efficiency -- i.e., the fraction of halo baryons that are converted into stars in galaxy hosting halos -- adopted is $f_\star = 0.05$.
The UV photon emissivity follows the Pop-II case described in \cite{Barkana:2004vb,Mesinger11}. 

In terms of X-ray heating, our fiducial model assumes that $\zeta_X = 2 \times 10^{56}$ X-ray photons are produced per solar mass incorporated in stars. The X-ray emission follows a power law spectrum (i.e., the specific luminosity is $L_\nu \propto \nu^{-\alpha_X}$ with spectral index $\alpha_X=1.2$ above a threshold frequency of $h \nu_{\rm \min,X}= 0.5$ keV. For further discussion regarding these parameters, we refer the reader to \cite{Mesinger11}. Finally, our 21cmFAST runs adopt the inhomogeneous recombination model of \cite{Sobacchi:2014rua}. Note that in \S \ref{sec:forecasts} we vary many of these parameters: $\zeta$, $T_{\rm vir}$, $f_\star$, and $\zeta_X$, in performing our Fisher matrix forecasts to account for parameter degeneracies.

\begin{figure}[htpb]
    \includegraphics[width=1.0\columnwidth]{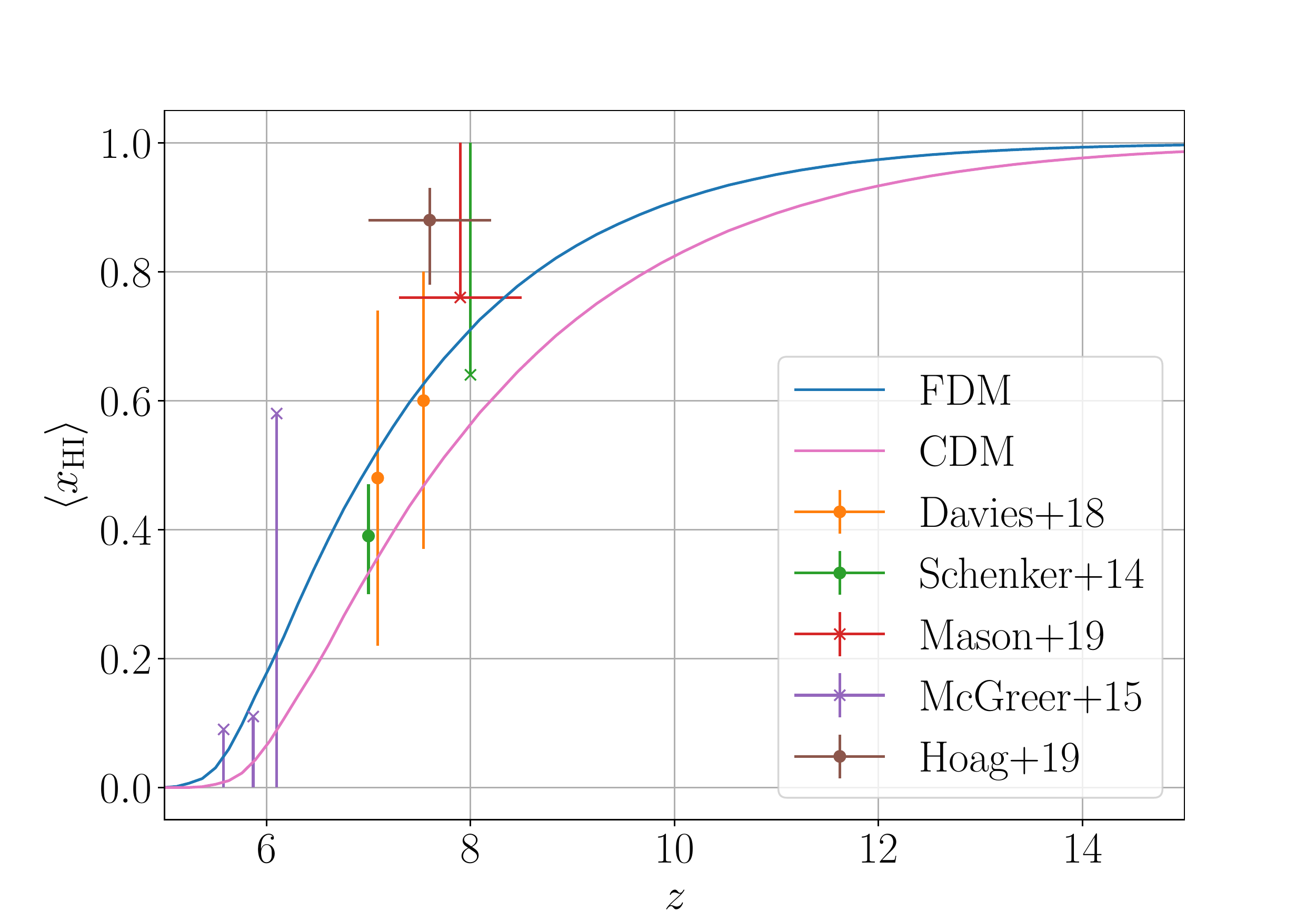}
    \caption{A comparison between the reionization history in our fiducial CDM and FDM models and current observational constraints. The magenta colored points show the volume-averaged neutral hydrogen fraction versus redshift in our CDM model with $\zeta=20$, $T_{\rm vir} = 2 \times 10^4$ K, while the blue points show the neutral fraction evolution for an FDM model with $m_{\rm FDM} = 1 \times 10^{-21}$ eV and the same $\zeta$, $T_{\rm vir}$. The data points show observational bounds on the reionizaion history, with $1-\sigma$ error estimates, compiled from the current literature. 
    }
    \label{fig:ionization_history}
\end{figure}

Fig~\ref{fig:ionization_history} compares our fiducial CDM model with current observational constraints on the ionization history, as inferred from: measurements of possible damping wing features in two $z \gtrsim 7$ quasars \citep{Davies:2018yfp}, observations of the redshift evolution of the fraction of photometrically selected Lyman-break galaxies that emit prominent Ly-$\alpha$ lines \citep{Schenker:2014tda,Mason:2019ixe,Hoag19},  and measurements of the dark pixel fraction in the Ly-$\alpha$ and Ly-$\beta$ forests towards background quasars \citep{McGreer:2014qwa}. We further compare these measurements with a fiducial FDM model (described in the next sub-section) of particle mass $m_{\rm FDM} = 1 \times 10^{-21}$ eV and identical $\zeta$, and $T_{\rm vir}$ to our CDM case. Overall, Fig~\ref{fig:ionization_history} illustrates broad consistency between each fiducial model and this compilation of current observational constraints. Note that our objective here is not to precisely match the current data, but merely to ensure that our baseline models are reasonable enough to reliably forecast the prospects for upcoming 21 cm observations. 

We can further compare these models with CMB measurements from the Planck satellite, which constrain the probability that CMB photons scatter off of free electrons produced during and after reionization. The Planck 2018 measurement of the electron scattering optical depth (specifically their combined TT, TE, EE, lowE, lensing + BAO constraint) is $\tau_e=0.0561 \pm 0.0071$, where the error bars are $1-\sigma$ confidence intervals \citep{Aghanim:2018eyx}. 
Our CDM and FDM models yield $\tau_e = 0.0675, 0.0580$, consistent with the Planck 2018 measurements at $1.6-\sigma$ and $0.27-\sigma$, respectively. 

Finally, it is worth commenting on how our fiducial models compare with the EDGES results \citep{Bowman:2018yin}, which suggest a deep 21 cm absorption signal starting at redshifts as high as $z\sim 20$. In our fiducial CDM model, the minimum absorption depth is reached at $z \sim 18$, while this is delayed until $z \sim 15$ in FDM. The redshift of the absorption dip in the CDM case is close to that of the EDGES measurement, although the depth and shape of this feature are quite different than observed (e.g. \citealt{Bowman:2018yin,Mirocha:2018cih}). Note, however, that our fiducial model assumes a star-formation efficiency of $f_\star = 0.05$ which is larger than suggested by UV luminosity function measurements and abundance matching constraints near $z \sim 8$ (see e.g. \citealt{Mirocha:2018cih,Lidz:2018fqo}). Adopting a lower star-formation efficiency would delay the onset of Cosmic Dawn and push the redshift of the absorption feature to lower redshifts.

\subsection{Modeling FDM with 21cmFAST}\label{sec:mod_fdm_21}

In order to model FDM with 21cmFAST, we adopt the approximation that FDM suppresses the initial power spectrum of density fluctuations on small scales, but we ignore the subsequent impact of FDM on the dynamics of structure formation. This is likely a good approximation for our application, essentially because the FDM Jeans mass drops with decreasing redshift
(see \citealt{Lidz:2018fqo} for further discussion and e.g. \citealt{Schive:2014dra,Li:2018kyk} for  simulation runs that follow the dynamical impact of so-called quantum pressure.)

With this simplification, we need only to modify the transfer function used in 21cmFAST in generating initial conditions. This in turn reduces the variance of the (linearly-extrapolated) density field at small smoothing scales, which suppresses the halo collapse fraction and thereby impacts the 21cmFAST excursion-set based modeling of the Cosmic Dawn and EoR. 

We use the FDM transfer function from \cite{Hu:2000ke}. 
Given the FDM particle mass in our model, $m_{\rm{FDM}}$,
this may be written as:
\beq\label{eq:tf_fdm}
\frac{P_{\rm{FDM}(k)}}{P_{\rm CDM}(k)} = \left[\frac{\rm{cos}(x^3(k))}{1+x^8(k)}\right]^2 \text{,}
\eeq
where $x(k) = 1.61 \left[m_{\rm{FDM}}/10^{-22} \rm{eV}\right]^{1/18} k/k_{\rm J, eq}$ and $k_{\rm J, eq}$ is the FDM Jeans wavenumber at matter-radiation equality, $k_{\rm J,eq} = 9.11\, {\rm Mpc}^{-1} \left[m_{\rm{FDM}}/10^{-22} \rm{eV}\right]^{1/2}$ \citep{Hu:2000ke}. This specifies the linear FDM power spectrum, $P_{\rm FDM}(k)$, in terms of the CDM one, $P_{\rm CDM}(k)$. 

The cutoff in the initial conditions, described by $k_{\rm J,eq}$, leads to a suppression in the halo mass function at small masses. It is useful to further describe this suppression by a characteristic halo mass scale, $M_{1/2}$. This is defined as the mass corresponding to the wavenumber, $k_{1/2}$, at which the linear FDM power spectrum is reduced by a factor of two relative to the CDM one with $M_{1/2} = \frac{4 \pi \rho_M}{3} \left(\pi/k_{1/2}\right)^3$. Here $\rho_M$ is the mean co-moving matter density. Numerically, the suppression mass is \citep{Hui:2016ltb}:
\begin{align}\label{eq:mhalf}
M_{1/2} &= 2.51 \times 10^9 M_\odot  \left(\frac{1 \times 10^{-21} {\rm eV}}{m_{\rm FDM}}\right)^{4/3} \nonumber \\
& \times \left(\frac{\Omega_m}{0.308}\right) \left(\frac{h}{0.678}\right)^{1/2}\text{.}
\end{align}
This mass scale is potentially larger than the characteristic host halo masses of the early generations of galaxies which formed during the Cosmic Dawn and the EoR, at least in CDM cosmological models. 
For example, the halo mass corresponding to a virial temperature of $T=10^4 {\rm K}$ --  at which point primordial gas can cool by atomic line emission, fragment, and form stars -- is $M=1.1 \times 10^8 M_\odot$ at $z=8$ (e.g. \citealt{Lidz:2018fqo}).\footnote{For further reference, the minimum galaxy hosting halo mass in our fiducial model with $T_{\rm vir} = 2 \times 10^4$ K is $M=3.1 \times 10^8 M_\odot$ at $z=8.$} 
Therefore the suppression of small-mass halos in FDM may delay the formation of the first galaxies relative to the CDM case, especially if galaxies are able to form efficiently in small mass CDM halos.

\section{The Impact of FDM on Cosmic Dawn and the EoR}\label{sec:qualitative_results}

\begin{figure*}
    \centering
    \includegraphics[scale=0.9]{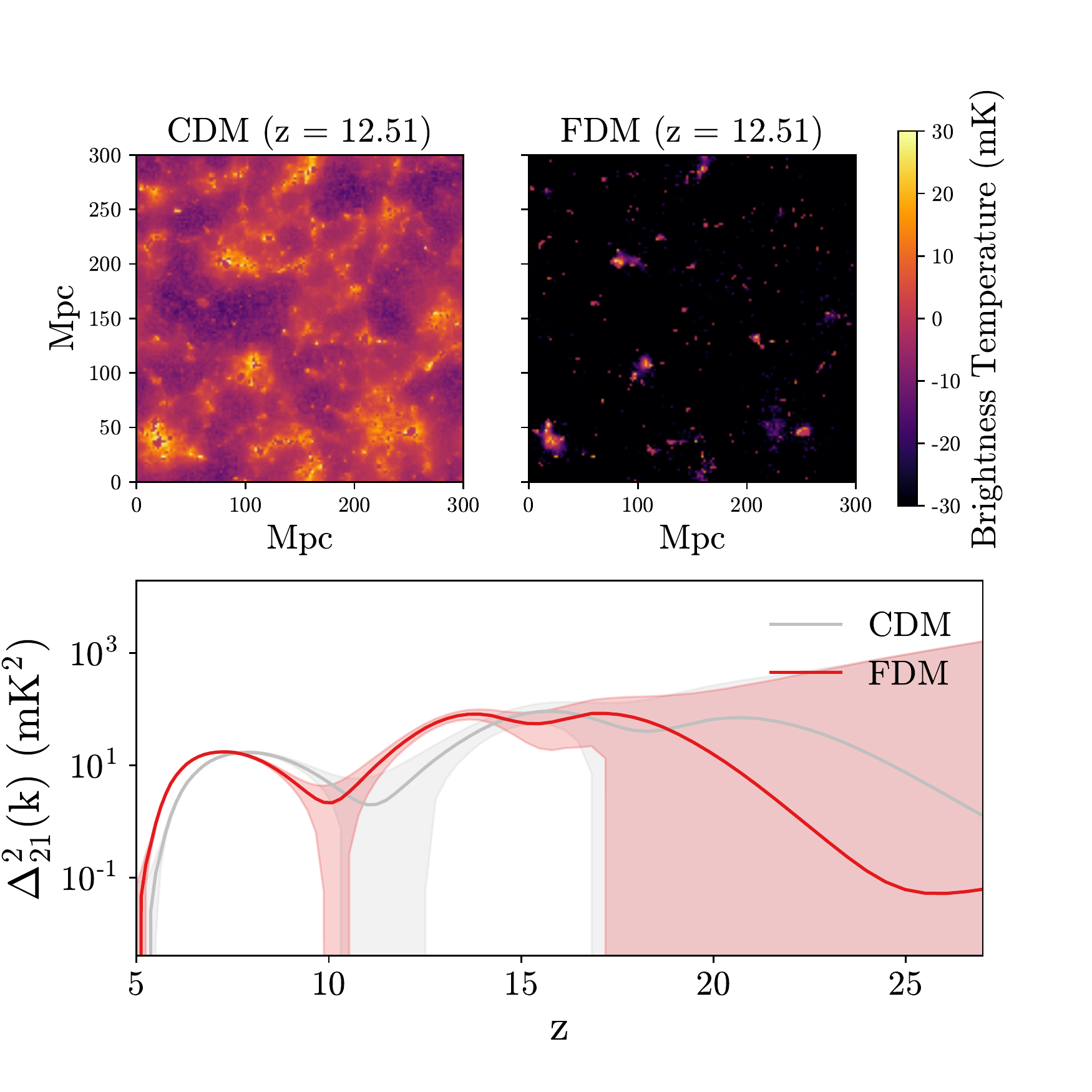}
    \caption{An overview of the results presented in this paper. {\it Top left}: A slice through the simulated 21 cm brightness temperature in our fiducial CDM model at $z=12.51$. The slice is 1.2 co-moving Mpc thick, and $300$ co-moving Mpc on a side. {\it Top right}: The corresponding slice though our fiducial FDM model, with $m_{\rm FDM} = 10^{-21} {\rm eV}$. The same random seeds and large-scale modes are adopted in the initial conditions for each simulation, and so a side-by-side comparison is warranted. While the 21 cm signal is observable in emission (red) across much of the simulation slice in the CDM model, most of the corresponding FDM model is still seen in absorption (blue). Note that although the 21 cm signal in the FDM case reaches brightness temperatures as low as $T_{21} = -140 {\rm mK}$, the color bar is symmetric around zero and saturates at $T_{21} = -30 {\rm mK}$. The partial absorption signal is a consequence of the delay in structure formation in FDM. {\it Bottom}: The full redshift evolution of the 21 cm power spectra at $k=0.2$ Mpc$^{-1}$ in CDM and FDM. The redshift evolution in FDM lags that in CDM, and the power spectra differ by large factors at several redshifts. The shaded regions give error bar forecasts for upcoming HERA observations (assuming the moderate foreground avoidance scenario, see \S \ref{sec:HERA}), illustrating that the two example models can be distinguished at high statistical significance.}
    \label{fig:summary}
\end{figure*}

\subsection{Summary}

Before going into more detail, Fig~\ref{fig:summary} provides a brief overview of what follows. The top panel of the figure contrasts example slices though our fiducial CDM and FDM models (with an FDM particle mass of $m_{\rm FDM} = 10^{-21} {\rm eV}$) at $z=12.51$. This particular redshift is chosen because it highlights some of the qualitative differences that arise. In the CDM model, early X-ray heating has already succeeded in raising the gas temperature much above the CMB temperature across a significant fraction of the simulation volume at this redshift, and so the 21 cm signal is observable in emission across much of the slice shown (Eq~\ref{eq:tbright}). For example, the average brightness temperature across the simulation volume is $-1.30$ mK and 
$38\%$ of the simulation volume has been heated above the CMB temperature at this redshift ($T_\gamma=37$ K).

Furthermore, reionization is underway in this model with a volume-averaged ionization fraction of $\avg{x_i} = 0.052$. In contrast, the suppression of small mass halos in the FDM model leads to much of the gas being seen in {\it absorption} relative to the CMB. Although early sources of Ly-$\alpha$ photons have managed to couple the spin temperature to the gas temperature globally in this model, much of the gas is still cooler than the CMB temperature and little of it is ionized. The average brightness temperature in the FDM model is $-56.8$ mK and just 
$1.1\%$ of the simulation volume has gas kinetic temperature above the CMB temperature. 
In FDM, X-ray emitting sources have only succeeded in forming around prominent overdensities and heated just relatively nearby gas above the CMB temperature (red regions), while most of the gas is cooler than the CMB (blue regions). Since the 21 cm brightness temperature is proportional to $1 - (T_\gamma/T_s)$ (Eq~\ref{eq:tbright}), the overall contrast in the 21 cm brightness temperature data cube is quite strong during stages of the Comic Dawn in which some of the gas is in absorption and some in emission. 

The bottom panel of Fig~\ref{fig:summary} gives a more quantitative summary, showing the full redshift evolution of the 21 cm power spectrum in CDM and FDM at an example wavenumber of $k=0.2$ Mpc$^{-1}$. As anticipated earlier, the FDM power spectrum evolution is delayed relative to the CDM one. One consequence of this is that the FDM power spectrum greatly exceeds the CDM one at certain redshifts. For example, near the redshift of the slices in the upper panel ($z=12.51$) the FDM power is enhanced relative to the CDM one by a factor of $\sim 5$. This occurs because much of the FDM volume at this redshift is in absorption, which leads to a larger contrast in the 21 cm brightness temperature than in CDM, where much of the gas is in emission. On the other hand, at some redshifts the CDM fluctuations exceed those in FDM: for example, at $z\sim 20-25$ the CDM power spectrum is much larger than in FDM 
because of the earlier Ly-a coupling in CDM. We discuss the different redshift evolution in these models further in what follows.
The shaded regions show the expected error bars from HERA assuming the moderate foreground contamination model from \citet{Pober:2013jna} (see \S \ref{sec:HERA}), demonstrating that these two example scenarios may be distinguished at high statistical significance, as we will quantify further subsequently. As discussed in \S \ref{sec:HERA}, the larger signal in the Cosmic Dawn epoch partly compensates for the increased thermal noise in the measurements, and so the model power spectra are potentially detectable at redshifts as large as $z \sim 15$ (see also e.g. \citealt{Ewall-Wice:2015uul}).

\begin{figure*}
    \centering
    \includegraphics[scale=0.7]{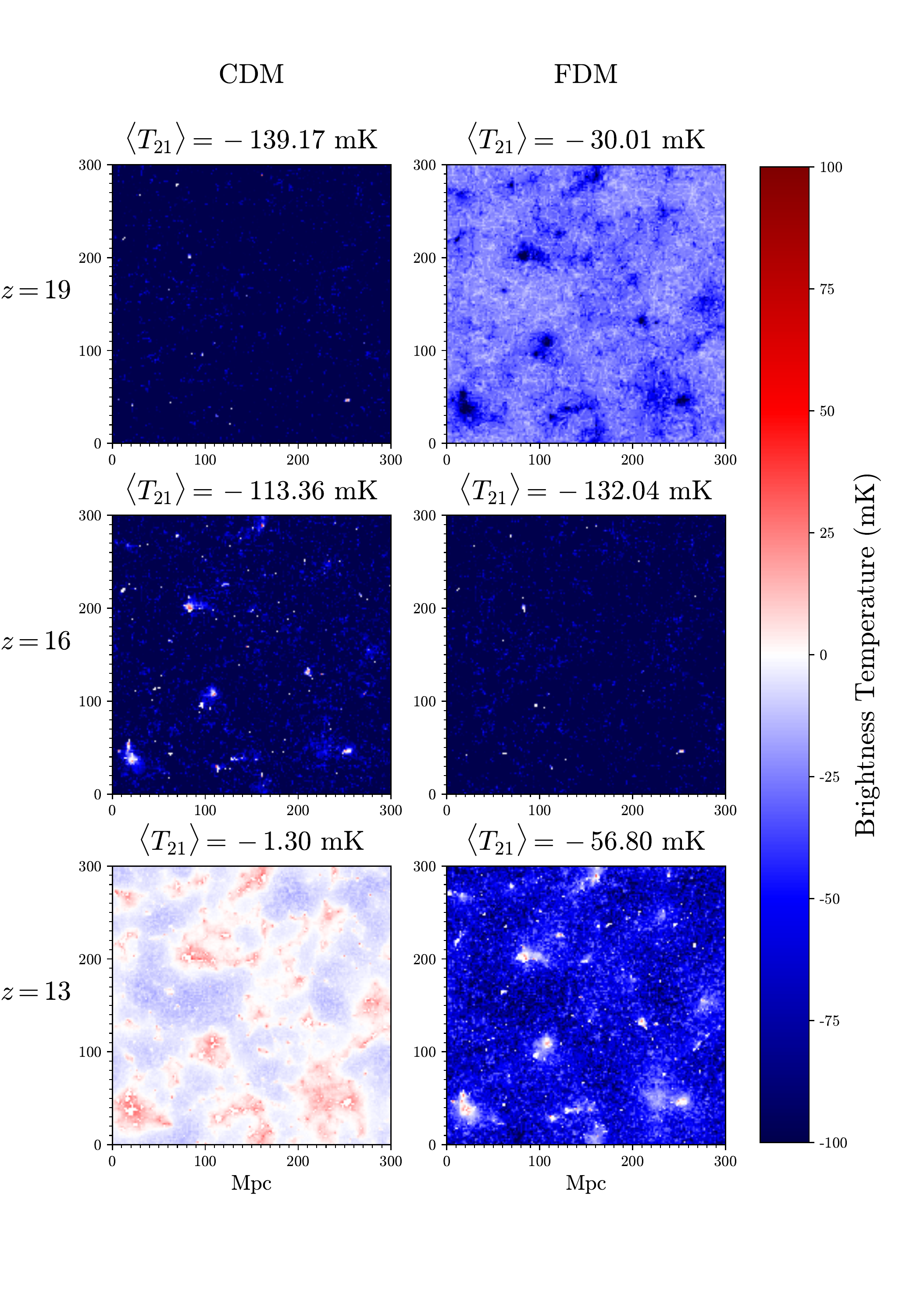}
    \caption{Brightness temperature evolution during Cosmic Dawn in our fiducial CDM and FDM models. {\it Left hand panels}: The CDM model at $z=18.69$, $z=15.80$, and $z=12.51$. {\it Right hand panels}: The FDM model at the same redshifts. The slice thickness and area are identical to those in Fig~\ref{fig:summary}, but here the color bar spans a larger range in brightness temperature in this figure. Note also that the bottom panel shows an identical redshift to the top panel of Fig~\ref{fig:summary}: they differ in appearance only because of the larger color bar range here.}
    \label{fig:slice_cdawn}
\end{figure*}

\subsection{Cosmic Dawn}

With this preview of the results to follow, we now systematically explore the full evolution of Cosmic Dawn and the EoR in our fiducial CDM and FDM scenarios.
The earliest phases of the Cosmic Dawn era involve the formation of the first stars, galaxies, and accreting black holes and their emission of ultraviolet (UV) photons. Some of these photons redshift into Lyman series resonances, and couple the spin temperature to the kinetic gas temperature via the Wouthuysen-Field effect \citep{Wouthuysen52,Field58}; this coupling requires on the order of one Lyman-alpha photon for every ten hydrogen atoms \citep{Chen:2003gc,Lidz:2018fqo}. This is expected to occur before the gas has been heated above the CMB temperature (e.g. \citealt{Pritchard2012})), and so the gas is cool and observable in 21 cm absorption during these early phases just after Ly-$\alpha$ coupling is achieved.  
In our fiducial CDM model, the upper left hand panel of Fig \ref{fig:slice_cdawn} shows that the 21 cm spin temperature is well-coupled to the gas temperature throughout most of the IGM volume in this simulation slice, at $z=18.69$. At the same redshift in FDM, the Wouthuysen-Field coupling is incomplete and so less of the simulation volume reaches the low brightness temperature seen in the CDM case. In the middle panel at $z=15.80$, the FDM model resembles the CDM scenario in the top panel (at $z=18.69$): the spin temperature is now well-coupled to the gas temperature across much of the simulation volume, and the FDM model shows a strong absorption signal. In the CDM model, early X-ray heating has started to boost the gas temperature to much above the CMB temperature in overdense regions, which are hence visible in emission (red regions). Finally, the bottom panel is identical to the top panel of the summary figure (Fig~\ref{fig:summary}), although we adopt a different color bar here. As discussed earlier, X-ray heating is well underway in the CDM case but the FDM model shows mostly absorption.

\begin{figure*}
    \centering
    \includegraphics[scale=.7]{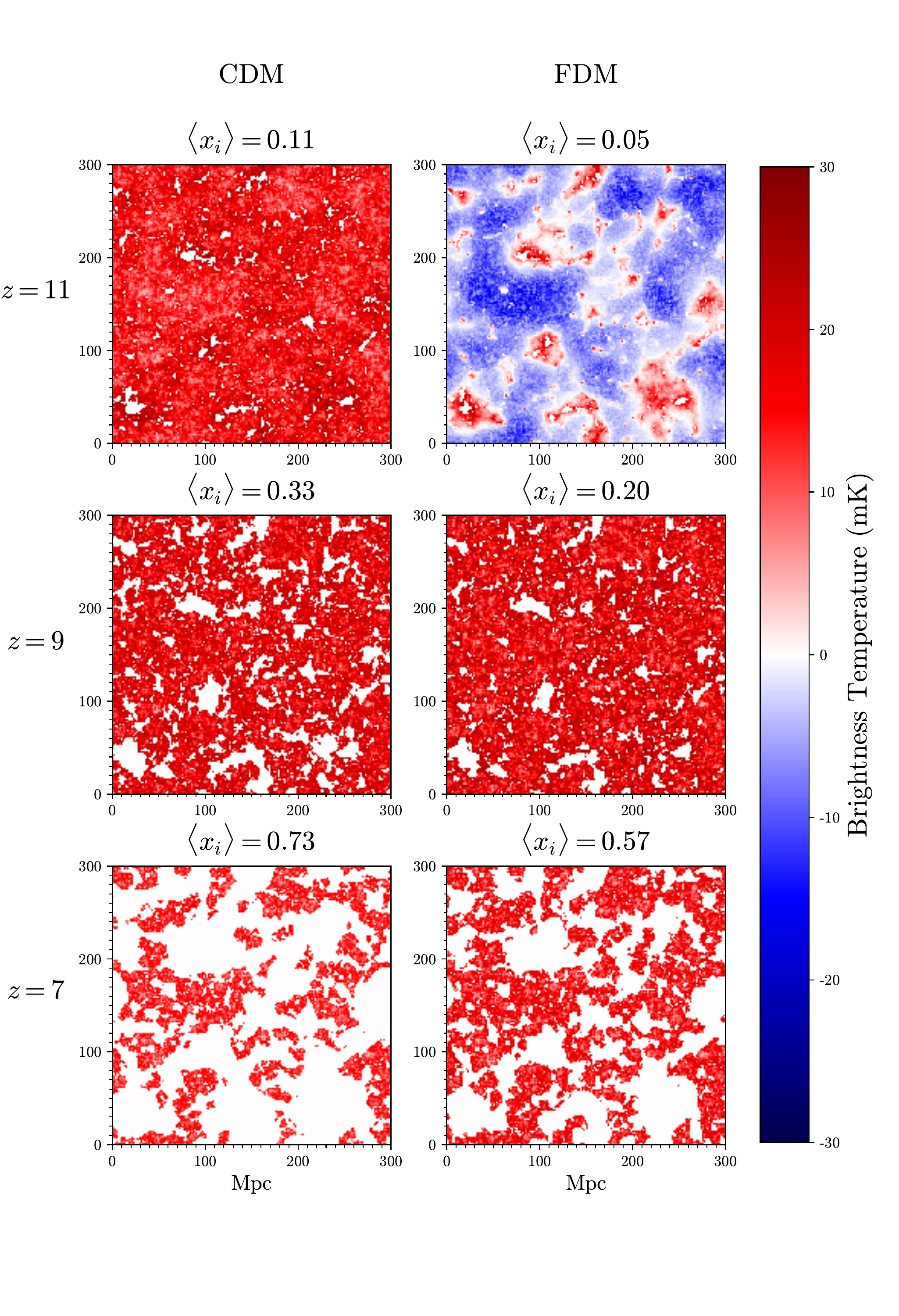}
    \caption{Brightness temperature evolution during the EoR in our fiducial CDM and FDM models. This is identical to Fig~\ref{fig:slice_cdawn}, but illustrates the evolution at redshifts of $z=11.00$, $z=8.65$, and $z=6.76$. The volume-averaged ionization fraction is given in each panel. The figure illustrates the delay in the EoR in FDM relative to CDM: the CDM model has a larger ionized fraction and somewhat bigger ionized bubbles.
    }
    \label{fig:slice_eor}
\end{figure*}

\subsection{EoR}

Fig~\ref{fig:slice_eor} displays slices through the simulation at slightly lower redshifts. The top panel shows each model at $z=11.00$: here FDM still shows a combination of 21 cm absorption/emission against the CMB, while the gas is everywhere heated above the CMB temperature in the CDM case. The middle and bottom panel illustrate how the EoR is more advanced in the case of CDM than FDM at redshifts $z=8.65$ and $z=6.76$. In terms of the volume-averaged ionization fraction, $\avg{x_i} = 0.109$ and $0.049$ at $z=11.00$, $\avg{x_i} = 0.326$ and $0.200$ at $z=8.65$, and $\avg{x_i} = 0.733$ and $0.567$ at $z=6.76$ in CDM and FDM, respectively. Naturally, the ionized regions are larger in CDM, mainly because the bubbles have had longer to grow and merge in this model. The absence of small halos in FDM also leads to larger ionized regions in FDM. These figures serve to qualitatively illustrate the delay in structure formation in FDM and the impact on the resulting 21 cm brightness temperature fluctuations.

\subsection{Power Spectra and the Impact of FDM on Spatial Structure}

\begin{figure}[htpb]
     \includegraphics[width=1.0 \columnwidth]{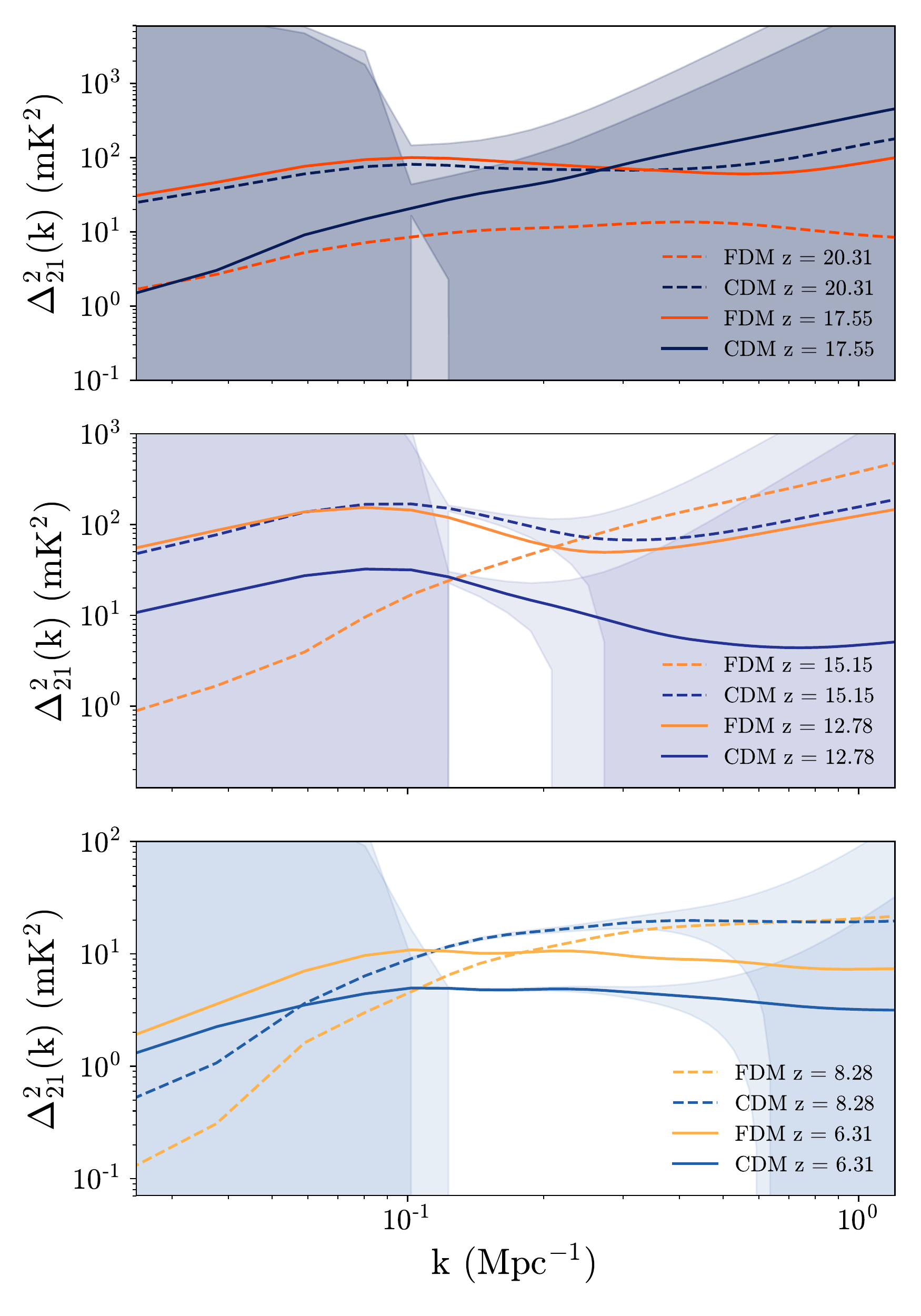}
     \caption{The 21 cm power spectra in the CDM and FDM models for six different example redshifts, two per panel. The solid lines show FDM models, while the dashed lines are for CDM.  {\it Top panel}: CDM and FDM power spectra at both $z=20.31$ and $z=17.55$. {\it Middle panel}: The power spectra in each model are displayed at $z=15.15$ and $z=12.78$. {\it Bottom panel}: CDM and FDM power spectra at $z=8.28$ and $z=6.31$. The shaded regions give error bar forecasts for upcoming HERA observations. At moderate scales, $k \sim 0.2$ Mpc$^{-1}$, the two models can be distinguished at high statistical significance across a range of redshifts.}
\label{fig:powerspectra}
\end{figure}

A more quantitative comparison is given in Fig~\ref{fig:powerspectra}.
In contrast to Fig~\ref{fig:summary}, which shows the power spectra as a function of redshift at one particular wavenumber, this figure presents the full scale dependence at six example redshifts, spanning the Cosmic Dawn and the EoR. These power spectra differ strikingly in shape and amplitude at many redshifts. For example, consider first the $z=15.15$ case in the middle panel of Fig~\ref{fig:powerspectra} (blue curves). Here, the CDM power spectrum exceeds that of FDM by a factor of $\sim 50$ on the largest scales shown (near $k \sim 0.03$ Mpc$^{-1}$, comparable to the fundamental mode of our simulation box). On the other hand, at higher wavenumber, $k \gtrsim 0.3$ Mpc$^{-1}$, the FDM model has more power than the CDM case at this redshift. 
The striking difference between the shape of the 21 cm power spectra in these models bodes well for distinguishing them with upcoming observations.

The top, highest redshift, panel compares the power spectra in both models at $z=17.55$ and $z=20.31$. The difference between the power spectra at these redshifts owes to the earlier Wouthuysen-Field effect coupling epoch in CDM. This leads to larger fluctuations in CDM across all scales shown at $z=20.31$, while FDM has larger fluctuations at $z=17.55$
for $k \lesssim 0.3$ Mpc$^{-1}$. The fluctuations are larger in FDM at the lower redshift because some regions of the universe in FDM are well-coupled and give deep 21 cm absorption, while other areas are close to the CMB temperature. This gives a larger contrast than the case of CDM where the spin temperature is well-coupled to the gas temperature across most of the simulation volume. Since the Wouthuysen-Field fluctuations are coherent on large scales, the excess power in FDM is concentrated at low $k$. 

The bottom panel of Fig~\ref{fig:powerspectra} shows the 21 cm power spectra during the EoR. Initially, as illustrated by the $z=8.09$ redshift case, the CDM fluctuations exceed the FDM ones at $k \lesssim 0.5$ Mpc$^{-1}$: this is a consequence of the larger ionized regions in the CDM model. However, by $z=6.31$ in the CDM model, the situation has reversed and the fluctuations are larger in FDM. This occurs because reionization is largely complete in CDM at this redshift, and the fluctuations are small since little neutral hydrogen remains, while reionization is less progressed in FDM.

Fig~\ref{fig:powerspectra} also includes HERA error bar forecasts in the moderate foreground removal scenario (see \S \ref{sec:HERA}). This illustrates that the models differ by more than the anticipated errors over a fairly broad range of scales and redshifts. Overall, the most valuable wavenumbers are in the intermediate range between roughly $k \sim 0.15-0.5$ Mpc$^{-1}$: the measurements on the largest scales are limited by foreground avoidance and sample variance, while the power at high-$k$ is swamped by thermal noise. As the thermal noise drops with increasing frequency (decreasing redshift), higher-$k$ modes become accessible. In terms of redshift, these forecasts suggest that HERA can discriminate between these models at high significance from the end of the EoR at $z \sim 5-6$ out to $z \sim 15$, with the error bars decreasing towards low redshift. Greater sensitivity would be required to detect the models at still higher redshifts (such as those in the top panel of Fig~\ref{fig:powerspectra}).

\begin{figure}[htpb]
    \centering
    \includegraphics[width=1.0 \columnwidth]{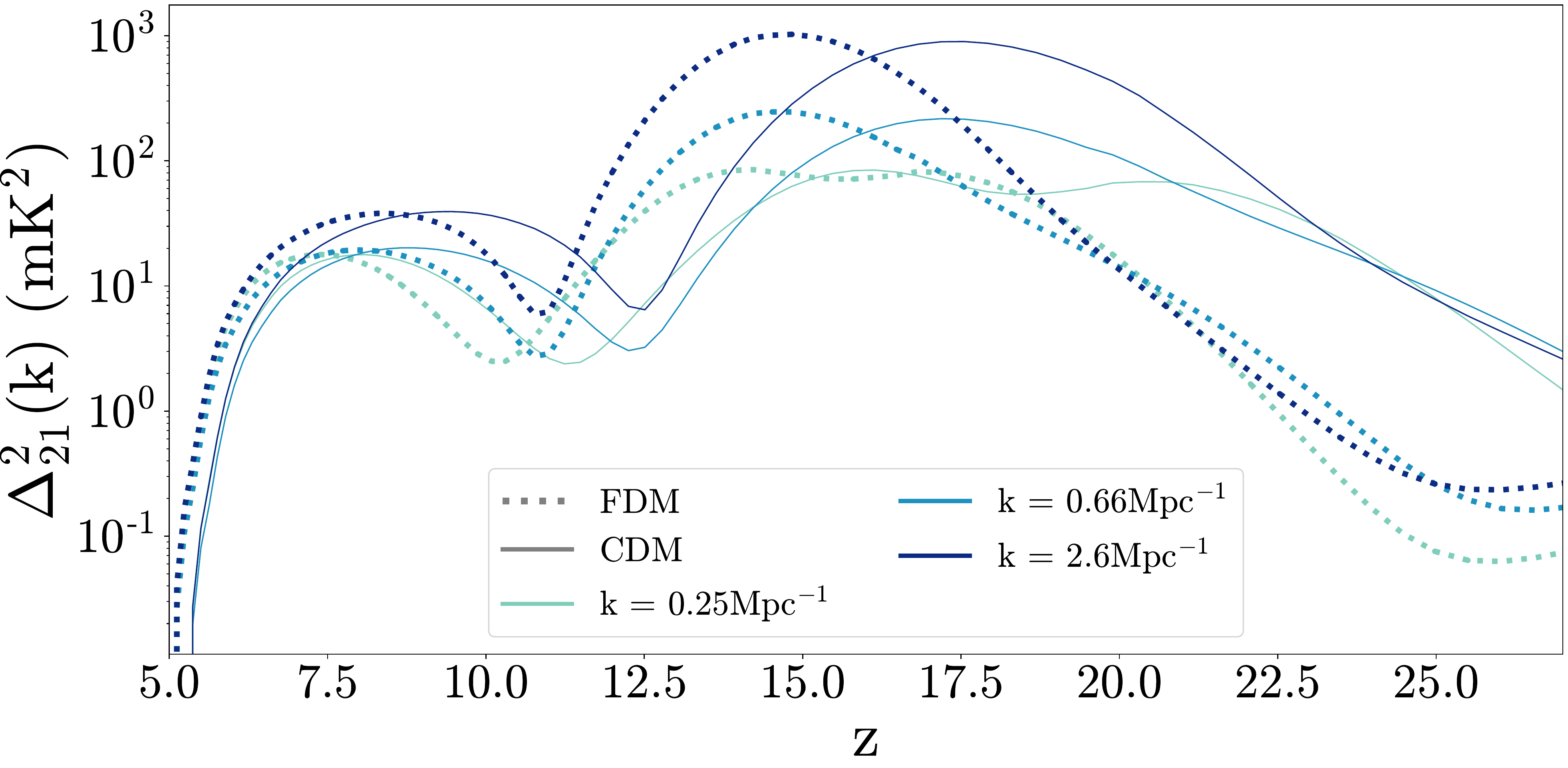}
    \caption{The full redshift evolution of the 21 cm power spectra at k = 0.25 Mpc$^{-1}$, k = 0.66 Mpc$^{-1}$, and k = 2.6 Mpc$^{-1}$ in CDM and FDM. This is similar to the bottom panel of Fig~\ref{fig:summary}, but here we show three different example wavenumbers.}
    \label{fig:PSevo}
\end{figure}

Fig~\ref{fig:PSevo} shows the redshift evolution in further detail for three example wavenumbers. This reinforces the trends seen in Fig~\ref{fig:summary} and Fig~\ref{fig:powerspectra} and illustrates the effects of the delay in structure formation in FDM at finer redshift evolution than in Fig~\ref{fig:powerspectra}.

\begin{figure*}
    \centering
    \includegraphics[scale=.5]{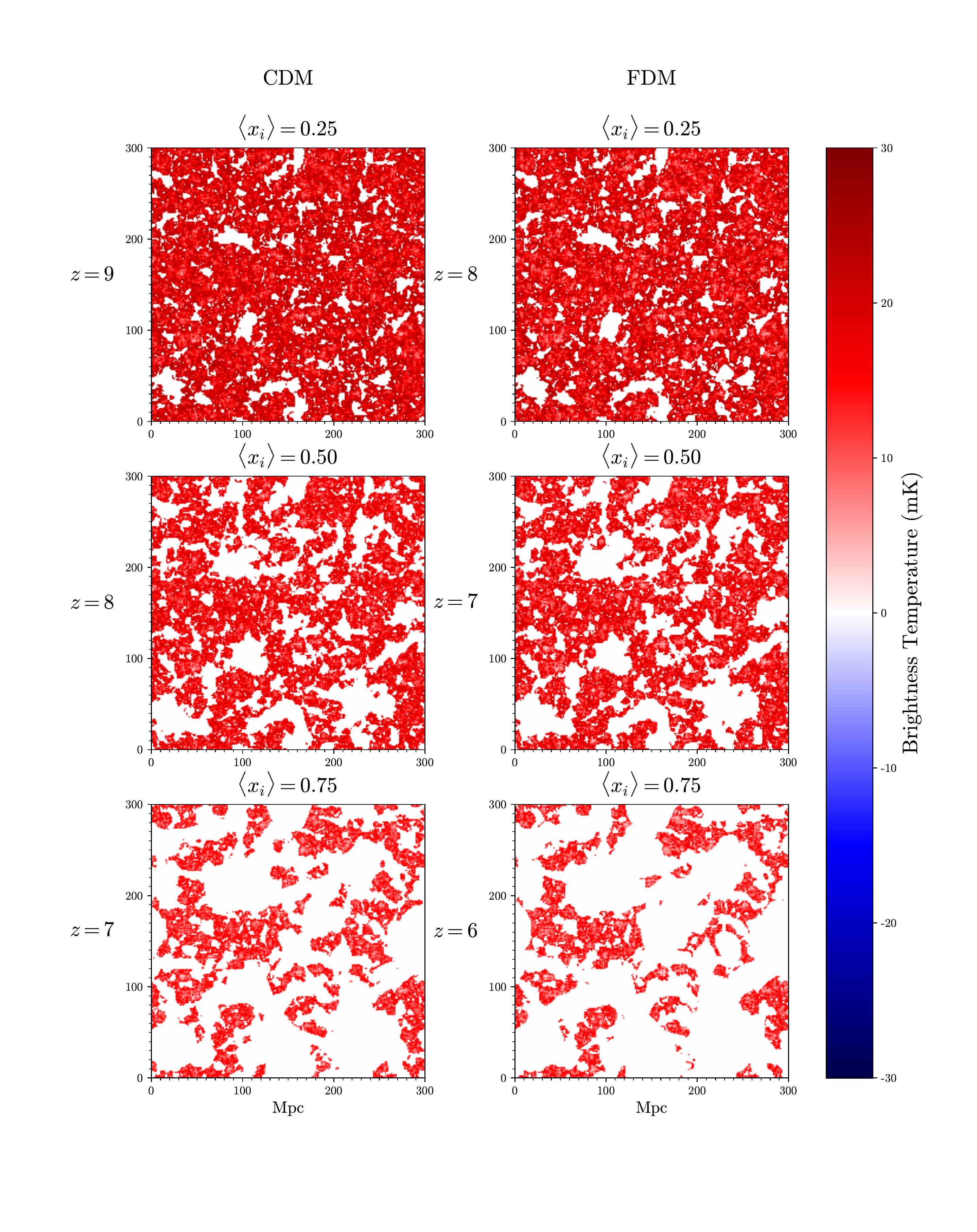}
    \caption{Brightness temperature in our fiducial CDM and FDM models at the same {\it stage} of the EoR, yet different redshifts. This figure is similar to Fig~\ref{fig:slice_cdawn} and \ref{fig:slice_eor} except the CDM and FDM models along each row are shown at fixed volume-averaged ionization fraction, $\avg{x_i}$, yet different redshifts. {\it Top row}: Each model is shown at $\avg{x_i}=0.25$ ( $z=9.24$ in CDM and $z=8.28$ for FDM). {\it Middle row}: Each model is given at $\avg{x_i}=0.50$. This occurs at $z=7.74$ in CDM and $z=7.08$ in FDM. {\it Bottom row}: Here the ionization fraction in each model is $\avg{x_i}=0.75$, at $z=6.76$ in CDM and $z=6.17$ in FDM.} 
    \label{fig:slice_same_stage}
\end{figure*}

While the different redshift evolution in CDM and FDM is interesting, note that the overall timing of reionization depends also on uncertain parameters such as the ionizing efficiency, $\zeta$, and the minimum virial temperature of galaxy hosting dark matter halos, $T_{\rm vir}$. It is also, therefore, interesting to contrast the 21 cm brightness temperature fluctuations in CDM and FDM models at the same {\it stage} of the Cosmic Dawn/EoR, yet different redshifts. This helps to understand how much of the effect of FDM is an overall delay in structure formation, and how much FDM impacts the overall spatial structure of Cosmic Dawn and the EoR. 

For example, Fig~\ref{fig:slice_same_stage} contrasts the CDM and FDM models during the EoR with slices drawn from each of $\avg{x_i}=0.25, 0.50$, and $0.75$. Although the effect is subtle, the ionized regions are slightly larger at fixed ionization fraction in FDM than in CDM (see also \citealt{Nebrin:2018vqt}). This result is seen because the size distribution of the ionized regions is sensitive to the {\it clustering} of the ionizing sources (e.g. \citealt{McQuinn:2006et}): the ionized regions are larger at a given stage of the EoR in cases where the ionizing sources lie in more massive -- and hence more highly biased -- dark matter halos. Since small mass halos are missing in FDM, this therefore leads to slightly larger ionized regions in FDM than CDM, at least for cases where the minimum galaxy host halo mass is smaller than the FDM suppression mass (Eq~\ref{eq:mhalf}). The larger ionized regions in FDM tend to boost the large-scale amplitude of the 21 cm power spectrum (at a given $\avg{x_i}$) in FDM relative to the CDM case. Hence, FDM modifies both the timing of reionization as well as the spatial structure of the 21 cm field. Although we do not illustrate it explicitly here, analogous effects also occur during the earlier Cosmic Dawn phases. That is, when we compare CDM and FDM at fixed average brightness temperature (yet differing redshifts), the greater source clustering in FDM enhances the large-scale 21 cm power spectra relative to CDM.

\section{HERA and Upcoming 21 cm Power Spectrum Measurements}\label{sec:HERA}

HERA is a radio interferometer, under development in the Karoo desert of South Africa, designed to detect the 21 cm signal from the EoR \citep{DeBoer:2016tnn}, and potentially the Cosmic Dawn \citep{Ewall-Wice:2015uul}, at high statistical significance. Readers already familiar with HERA may wish to skip to the second to last paragraph of this section.
When complete, HERA will consist of 350 antenna dishes, each 14 meters in diameter, with 320 of these in a close-packed hexagonal configuration, along with 30 outrigger antennas at longer baselines.  The close-packed hexagonal configuration provides a highly redundant sampling of baselines, with many identical copies of the same antenna separations; this helps achieve high 21 cm power spectrum sensitivity while facilitating instrumental calibration \citep{Dillon:2016ljv}. Ultimately, the instrument will observe a broad frequency range from 50-225 MHz, corresponding to redshifted 21 cm radiation from $z \sim 5-27$. The array always points towards the zenith, but the interferometer operates as a drift-scan telescope, accumulating sky coverage as the sky revolves overhead owing to the rotation of the Earth.

In order to quantify the prospects for HERA measurements of the 21 cm power spectrum, and its ability to discriminate between CDM and FDM models, we make use of the open-source Python package \textsc{21cmSense} \citep{Pober13,Pober:2013jna}. In brief (see e.g. \citealt{Pober:2013jna,Liu:2019awk} for more details), the \textsc{21cmSense} code accounts for the detailed layout of the HERA antennas, gridding the measurements into cells in the $uv$ plane, where $u$ and $v$ describe the physical separation between a pair of antenna dishes in units of observed wavelength. The size of each $uv$ cell is set by the diameter of the HERA dishes and is of order $D/\lambda_{obs}$ on a side, where $D$ is the dish diameter. Further, each cell has a width in $\eta$, where $\eta$ is the Fourier counterpart to frequency, set by the frequency bandwidth of the measurement, $B$.
The code calculates the observing time per $uv$ cell accounting for the rotation of the Earth which causes baselines to move across $uv$ cells over the course of a day. The interferometric $uv$ cells sample Fourier modes of transverse wavenumber, $\k_\perp = 2 \pi {\bf b}/X(z)$, where ${\bf b}$ is a baseline vector in the $uv$ plane and $X(z)$ is the co-moving angular diameter distance to the 21 cm redshift at the central frequency across the bandwidth of interest.  The $\eta$ dimension maps to the line-of-sight wavenumber component, $k_\parallel$ (see, e.g., Eq 40 of \citealt{Liu:2019awk}). 

After determining the total observing time $t(\k)$ for each $u,v,\eta$ cell, the variance of the power spectrum estimate in a cell is given by \citep{Pober:2013jna,Ewall-Wice:2015uul}:
\begin{equation}\label{eq:power_variance}
\sigma_P^2(\k) = \left[X^2 Y \frac{\Omega^\prime}{2 t(\k)} T_{\rm sys}^2 + P_{21}(k)\right]^2,
\end{equation}
where $X$ is the co-moving angular diameter distance, and $Y$ is a redshift dependent factor that converts between frequency intervals and co-moving line-of sight distance (e.g. Eq 41 of \citealt{Liu:2019awk}). The quantity $\Omega^\prime$ is a factor related to the solid angle of the primary beam. Specifically, it is the integral of the primary beam squared over solid angle divided by the solid-angle integral of the primary beam \citep{Parsons14}, while $T_{\rm sys}$ is the sum of the HERA receiver temperature and the sky temperature.\footnote{The \textsc{21cmSense} codes assumes a receiver temperature of $100$ K and a sky temperature of $T=60\, \mathrm{K} \, (\nu/300 \mathrm{MHz})^{-2.55}$.} The $P_{21}(k)$ term accounts for sample variance under the Gaussian error approximation, and is determined by our 21cmFAST model under consideration.  
Finally, in order to estimate the variance across different $k$-bins, \textsc{21cmSense} adds the errors from Eq \ref{eq:power_variance} over contributing $\k$ cells in inverse quadrature.

In forecasting the HERA sensitivity, we assume 1080 total hours of observing time, and that power spectra are simultaneously measured from individual bandwidths of frequency extent $8$ MHz across the entire observing range from $z \sim 5-27$. It may be unrealistic to assume that simultaneous measurements are feasible across this large frequency band \citep{Ewall-Wice:2015uul}, but this should nevertheless provide a useful, if optimistic, forecast. Note also that redshifted 21 cm observations between $z \sim 12-15$ lie within the FM radio band, where radio-frequency interference mitigation may be especially challenging \citep{Ewall-Wice:2015uul}. We quantify how the expected signal to noise ratio varies with redshift, and so determine which observed frequencies across this broad range may be most valuable.

Foreground contamination is a serious concern for redshifted 21 cm measurements (see e.g. the review by \citealt{Liu:2019awk}): the foreground emission from sources including galactic synchrotron radiation, free-free emission, and extra-galactic point sources is many orders of magnitude brighter than the redshifted 21 cm signal. Nevertheless, the foregrounds are expected to be spectrally smooth while the 21 cm signal has a great deal of spectral structure, and this distinction holds promise for separating the signal from the foregrounds. Specifically, spectrally smooth foregrounds will strongly contaminate low $k_\parallel$ modes, while higher $k_\parallel$ modes may be robustly measurable. Accounting, however, for the frequency dependence of the instrumental response leads to a mode-mixing effect in which some high $k_\parallel$ modes are also corrupted by foregrounds. Still, the corrupted modes should mostly occupy a wedge-shaped region in the $k_\parallel-k_\perp$ plane referred to as ``the foreground wedge'' in \citet{Pober_2014}. A promising strategy is then to simply excise Fourier modes within the foreground wedge and make use only of regions in $\k$-space outside of this wedge. In practice, the precise form of the foreground wedge is uncertain owing to (see \citealt{Liu:2019awk} for further discussion): the unknown $k_\parallel$ dependence of the foreground emission, the impact of calibration errors, and instrumental artifacts, with some effects potentially leaking power outside of the wedge entirely.

To roughly quantify the uncertain impact of foreground contamination, we follow the three separate treatments of the foreground wedge discussed in \citet{Pober:2013jna} and included in the \textsc{21cmSense} code, termed the ``pessimistic, moderate, and optimistic'' foreground scenarios. These cases are only briefly summarized here; we refer the reader to the original paper for further details. In these scenarios, the ``horizon wedge'' describes a line with $k_\parallel=C k_\perp$ (where $C$ is a redshift dependent number), below which a population of spatially unclustered radio sources at the horizon, with a frequency-independent emission spectrum, will contaminate measurements (see e.g. Eq 166 in \citealt{Liu:2019awk} and the associated discussion). Above this line, such sources produce no contamination. In \citet{Pober:2013jna}'s moderate case, the wedge is assumed to extend to $\Delta k_\parallel = 0.1$ $h$Mpc$^{-1}$ beyond the horizon wedge limit. In the optimistic case, the angular scale defining the wedge (which determines $C$) is assumed to be set by the FWHM of the primary beam of HERA, rather than the horizon scale. Finally, in the pessimistic case the horizon wedge is assumed, but only instantaneously redundant baselines, or baselines that measure the same Fourier component
of the sky brightness distribution (\citet{10.1093/mnras/stt1902}), are added coherently.

The resulting power spectrum sensitivities in the moderate foreground case are shown at several example redshifts in the figures of the previous section (Figs~\ref{fig:summary} and \ref{fig:powerspectra}), illustrating the usual high significance forecasts for HERA power spectrum measurements (e.g. \citealt{DeBoer:2016tnn}).

\begin{figure}[htpb]
    \includegraphics[width=\columnwidth]{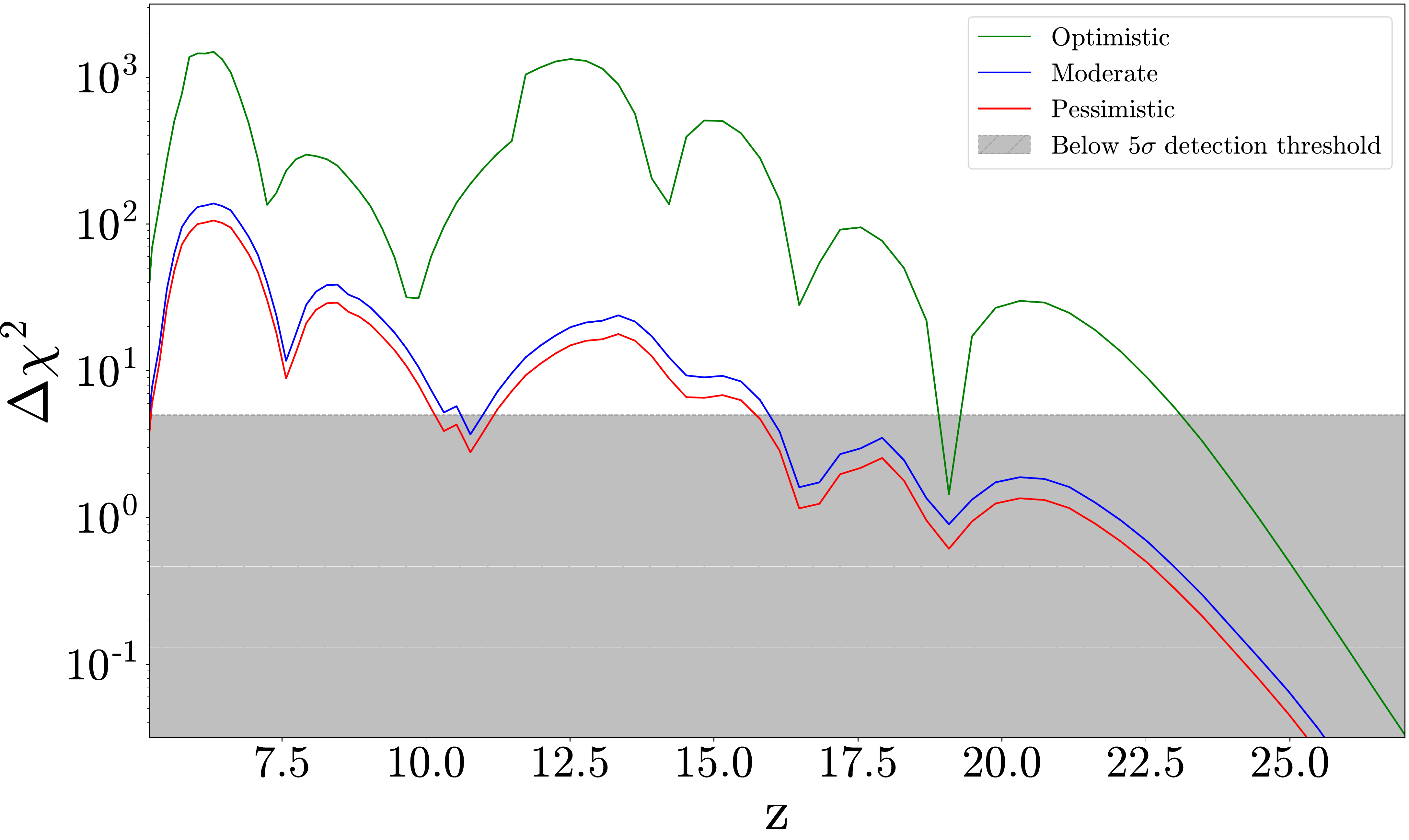}
    \caption{
The statistical significance at which HERA can distinguish between our fiducial CDM and FDM models as a function of redshift. The particle mass in the FDM model is $m_{\rm FDM} = 10^{-21}$ eV. The green, blue, and red curves show the optimistic, moderate, and pessimistic foreground wedge models (see text), respectively. In the absence of parameter degeneracies, the two models may be discriminated at $\geq 5-\sigma$ significance for each foreground treatment -- as shown by regions above the grey band -- across a range of redshifts. 
}\label{fig:chisq_cdm_fdm_fiducial}
\end{figure}

To gain further insight into the prospects for discriminating between CDM and FDM with HERA 21 cm power spectrum measurements, we calculate $\Delta \chi^2$ between our two fiducial models as a function of redshift, assuming that the CDM case is the true underlying model. Encouragingly, as illustrated in Fig~\ref{fig:chisq_cdm_fdm_fiducial}, these two models may be discriminated at high statistical significance across a range of redshifts for all foreground contamination scenarios. Formally, in the optimistic case the two models may be distinguished at more than $100-\sigma$, with this level of discriminating power achievable in multiple independent redshift bins. This is encouraging, especially considering that our fiducial FDM model has a particle mass of $m_{\rm FDM} = 10^{-21}$ eV, fairly comparable to current limits from the Ly-$\alpha$ forest \citep{Irsic:2017yje}, although more stringent limits were found recently from Ly-$\alpha$ data by \cite{Rogers:2020ltq}. In any case, HERA measurements may provide an independent and potentially powerful constraint on FDM, although this statement depends somewhat on the impact of degeneracies with astrophysical parameters, as studied in the next section.

Fig~\ref{fig:chisq_cdm_fdm_fiducial} also reveals interesting trends in the constraining power versus redshift. In all cases, the strongest differences between the two models (relative to the HERA error bars) occurs towards the end of reionization, but $\Delta \chi^2$ also shows prominent peaks near $z \sim 12.5$. This redshift dependence arises because the sky background, dominated by galactic synchrotron emission, scales as $\nu^{-2.6}$ and so the noise power spectrum -- which is quadratic in the sky temperature -- scales strongly with redshift. On the other hand, the signal power spectrum and the difference between models is actually {\it larger} at high redshift during the Cosmic Dawn (see \S \ref{sec:qualitative_results} and Fig~\ref{fig:powerspectra}), which partly compensates for the enhanced noise. Intuition for the bumpy structure in Fig~\ref{fig:chisq_cdm_fdm_fiducial} can be gleaned from Fig~\ref{fig:summary}: FDM is mostly a delayed version of the CDM case, and at some redshifts the power spectra differ greatly in magnitude while at others they happen to be nearer in amplitude. These differences lead to corresponding structure in the $\Delta \chi^2$ curves although the exact location of these bumps will change for different reionization models.

\section{Fisher Matrix Forecasts}\label{sec:forecasts}

In order to make more quantitative forecasts, however, we need to account for parameter degeneracies. We accomplish this using the Fisher matrix formalism. 
Specifically, we consider a five-dimensional parameter space described by a vector ${\bf q}$ with components $(q_{\zeta}, q_{\rm Tvir}, q_{\rm f\star}, q_{\rm \zeta X}, q_{mFDM})$. The parameters here describe the fractional difference between each quantity and our fiducial model (e.g. \citealt{Ewall-Wice:2015uul}): e.g. $q_{\rm Tvir} = (T_{\rm vir} - T_{\rm vir,fid})/T_{\rm vir,fid}$.  As discussed previously in \S \ref{sec:method}, $\zeta$ is an ionizing efficiency parameter, $T_{\rm vir}$ is the minimum virial temperature of galaxy hosting dark matter halos, $f_\star$ is the star formation efficiency, $\zeta_X$ is an X-ray heating efficiency parameter, and $m_{\rm FDM}$ is the FDM particle mass. For simplicity, we assume the astrophysical parameters are redshift independent; we comment further on this assumption in what follows. As discussed earlier, our fiducial parameter set is: $(\zeta=20, T_{\rm vir} = 2 \times 10^4 {\rm K}, f_\star = 0.05, \zeta_X=2 \times 10^{56} \, M_\odot^{-1}, m_{\rm FDM} = 10^{-21} {\rm eV})$. Note that the fiducial model here is an FDM one, rather than a CDM case, since this facilitates the Fisher matrix computations.\footnote{First, a fiducial CDM case would involve an effectively infinite FDM mass. This problem could be avoided by adopting the inverse FDM mass as the model parameter rather than the mass itself. However,
this leads to asymmetric errors and violates the Fisher formalism's assumption of a quadratic expansion in the log-likelihood around the fiducial parameter values. Although techniques have been proposed in the context of warm dark matter models to circumvent these issues \citep{Markovic11}, we avoid them here by simply assuming an FDM case as our fiducial model. This is sufficient for our goals of understanding the impact of parameter degeneracies and the overall ability of the HERA data to constrain FDM mass.} 

The Fisher matrix may be written as:
\beq\label{eq:fisher}
F_{ij} =\sum_{k,z} \frac{\partial \Delta^2_{21}(k,z)}{\partial q_i} \frac{\partial \Delta^2_{21}(k,z)}{\partial q_j} \frac{1}{{\rm var}[\Delta^2_{21}[k,z]]} \text{,}
\eeq
where the sum runs over the full range of redshift and wavenumber bins, and the power spectrum variance is computed using \textsc{21cmSense} as described in the previous section. The wavenumber bins and redshift bins, separated by the $B=8$ MHz bandwidth of each power spectrum measurement, are approximated as independent. The resulting parameter constraint forecasts are obtained by computing the inverse of the Fisher matrix. We compute the derivatives with respect to the various parameters in Eq~\ref{eq:fisher} using two-sided numerical derivatives with a step-size of $5\%$ in each parameter. We find nearly identical results using one-sided derivatives with the same step size. 

\begin{figure}[htpb]
    \centering
     \includegraphics[width=1.0\columnwidth]{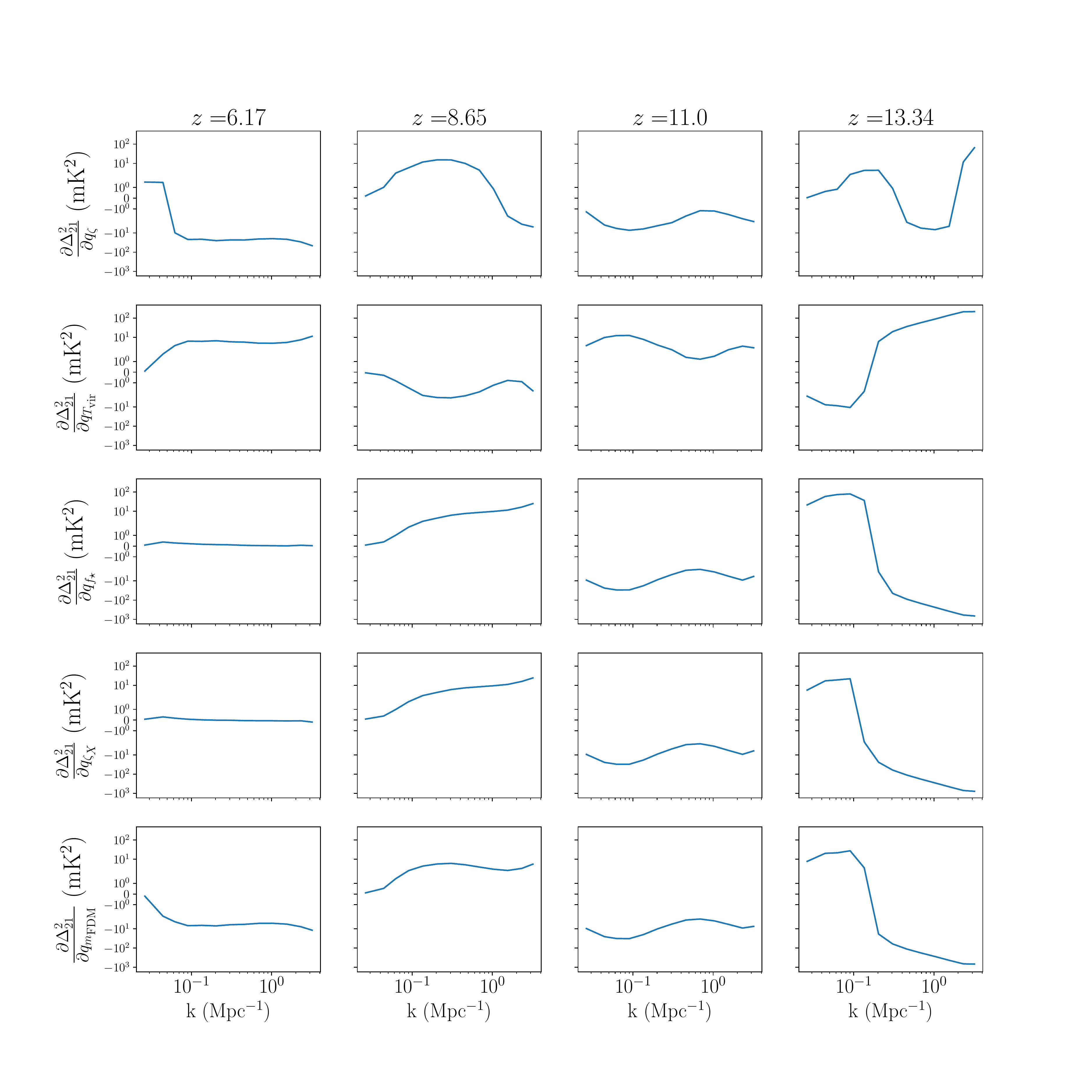}
    \caption{Derivatives of the 21 cm power spectrum with respect to various parameters.  These derivatives enter into the Fisher matrix computations of Eq~\ref{eq:fisher} and show how the power spectrum depends on parameters. For illustration, the results are shown at several example redshifts as a function of $k$.}
\label{fig:fisher_matrix_derivatives}
\end{figure}

\begin{figure}[htpb]
    \centering
     \includegraphics[width=1.0\columnwidth]{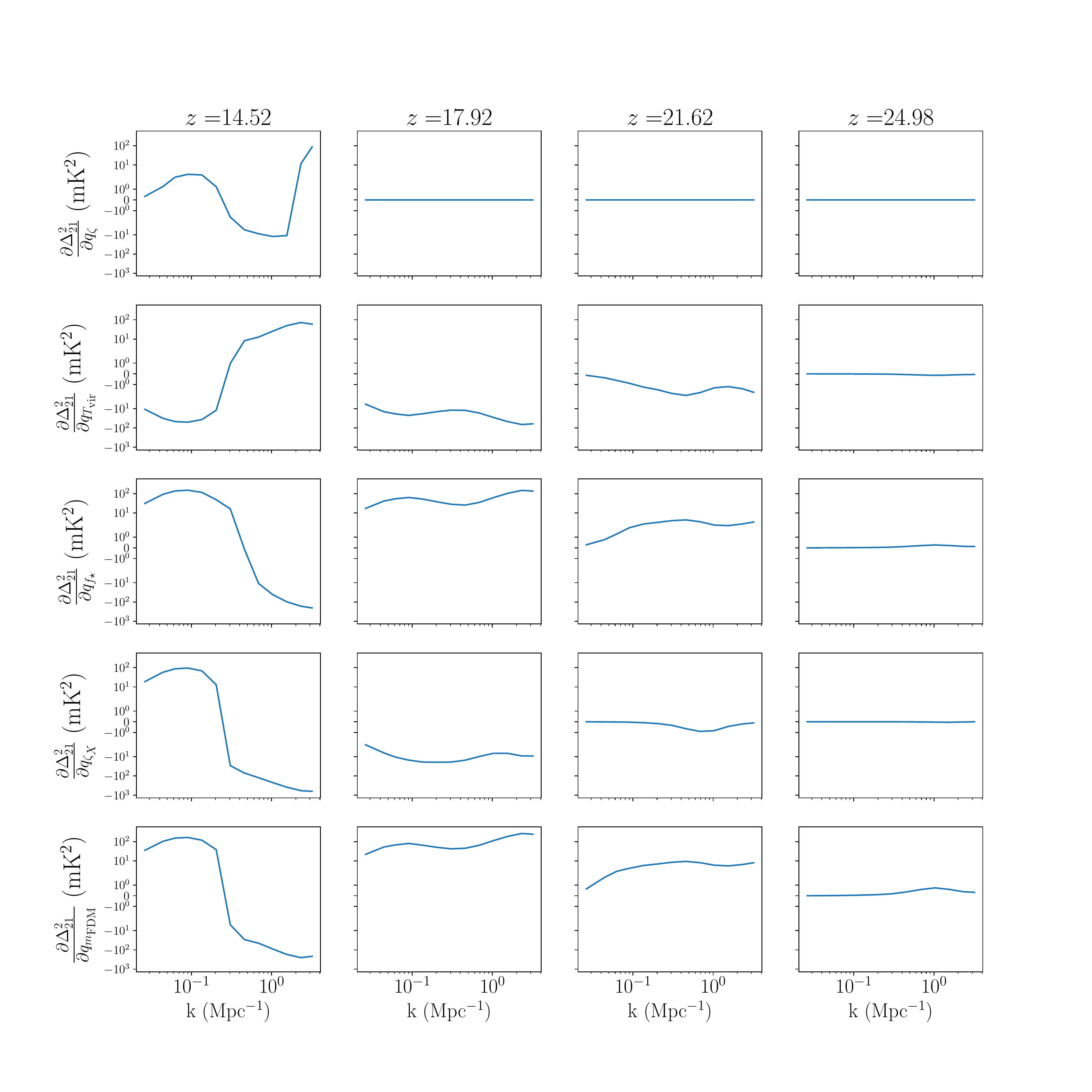}
    \caption{ Identical to Figure~\ref{fig:fisher_matrix_derivatives} but at higher redshifts.}
\label{fig:fisher_matrix_derivatives2}
\end{figure}

It is instructive to first examine the derivatives with respect to the five parameters. Fig~\ref{fig:fisher_matrix_derivatives} shows derivatives at several example redshifts as a function of wavenumber. We focus our attention on how the derivatives with respect to FDM mass compare with the other parameter derivatives. During the EoR (at $z \lesssim 10$ in our fiducial model), the derivatives with respect to FDM mass and $q_\zeta$ share the same sign and have a similar scale dependence. This is expected because increasing the FDM particle mass reduces the halo abundance suppression effect from FDM. This lessens the resulting delay in structure formation and in the timing of the EoR, while increasing the ionizing efficiency $\zeta$ has a similar effect. In other words, we expect the error bars for these two parameters to show a negative correlation since one can compensate for the reduced suppression from raising the FDM mass by decreasing the ionizing efficiency. In practice, however, allowing further parameters to vary impacts this degeneracy direction as discussed further below.

The degeneracy with $\zeta$ can in principle be broken by observations at higher redshift. During the Cosmic Dawn, when only a very small fraction of the IGM volume is ionized, the ionizing efficiency is not by itself an important parameter. In these earlier epochs, the X-ray heating and star-formation efficiency parameters are instead important. Although we generally expect there to be some relationship between the ionizing and star-formation efficiencies, we treat these as independent parameters since the ionizing efficiency depends additionally on the escape fraction of ionizing photons, for example. At high redshifts, the star-formation efficiency in our model plays a key role in determining the onset of Wouthuysen-Field coupling. Therefore, at higher redshifts, the derivatives with respect to FDM should be compared to those with respect to $q_{f \star}$ and $q_{\zeta x}$, while there is a different and much weaker dependence on $q_{\zeta}$ (as illustrated by the two highest redshift panels in Fig~\ref{fig:fisher_matrix_derivatives}). Furthermore, $q_{f \star}$ impacts mostly higher redshifts than $q_{\zeta x}$ since the  Wouthuysen-Field coupling precedes X-ray heating in our fiducial model. This is shown explicitly in Fig~\ref{fig:fisher_matrix_derivatives2}. The differing trends with redshift imply that HERA and other 21 cm surveys can help break degeneracies between FDM mass and astrophysical parameters by measuring the full redshift evolution of the signal, especially if this can be done over a relatively broad range in wavenumber. The one caveat here is that we have assumed the various astrophysical parameters are redshift independent: allowing redshift evolution in these parameters would naturally weaken our forecasts on FDM mass. We suspect, however, that this is not a strong effect for plausible smooth and monotonic redshift variations in these parameters. 

The other important astrophysical parameter at play is $T_{\rm vir}$ which sets the minimum mass of galaxy hosting dark matter halos in our model. Increasing the virial temperature suppresses the abundance of galaxy hosting halos. This effect resembles decreasing the FDM mass, and so we anticipate positively correlated errors on virial temperature and FDM mass. This degeneracy is reflected by the opposite signs, yet similar shape, of the derivatives with respect to $q_{\rm Tvir}$ and $q_{\rm mFDM}$ in Fig~\ref{fig:fisher_matrix_derivatives} and Fig~\ref{fig:fisher_matrix_derivatives2}. Although this is an important degeneracy, note that the FDM suppression mass in our fiducial model is almost an order of magnitude larger than the mass associated with our fiducial value of the virial temperature (see Eq~\ref{eq:mhalf} and the discussion in \S \ref{sec:mod_fdm_21}). Therefore, sharp constraints on FDM mass are still expected in our fiducial model.

\begin{figure}[htpb]
    \centering
\includegraphics[width=1.0\columnwidth]{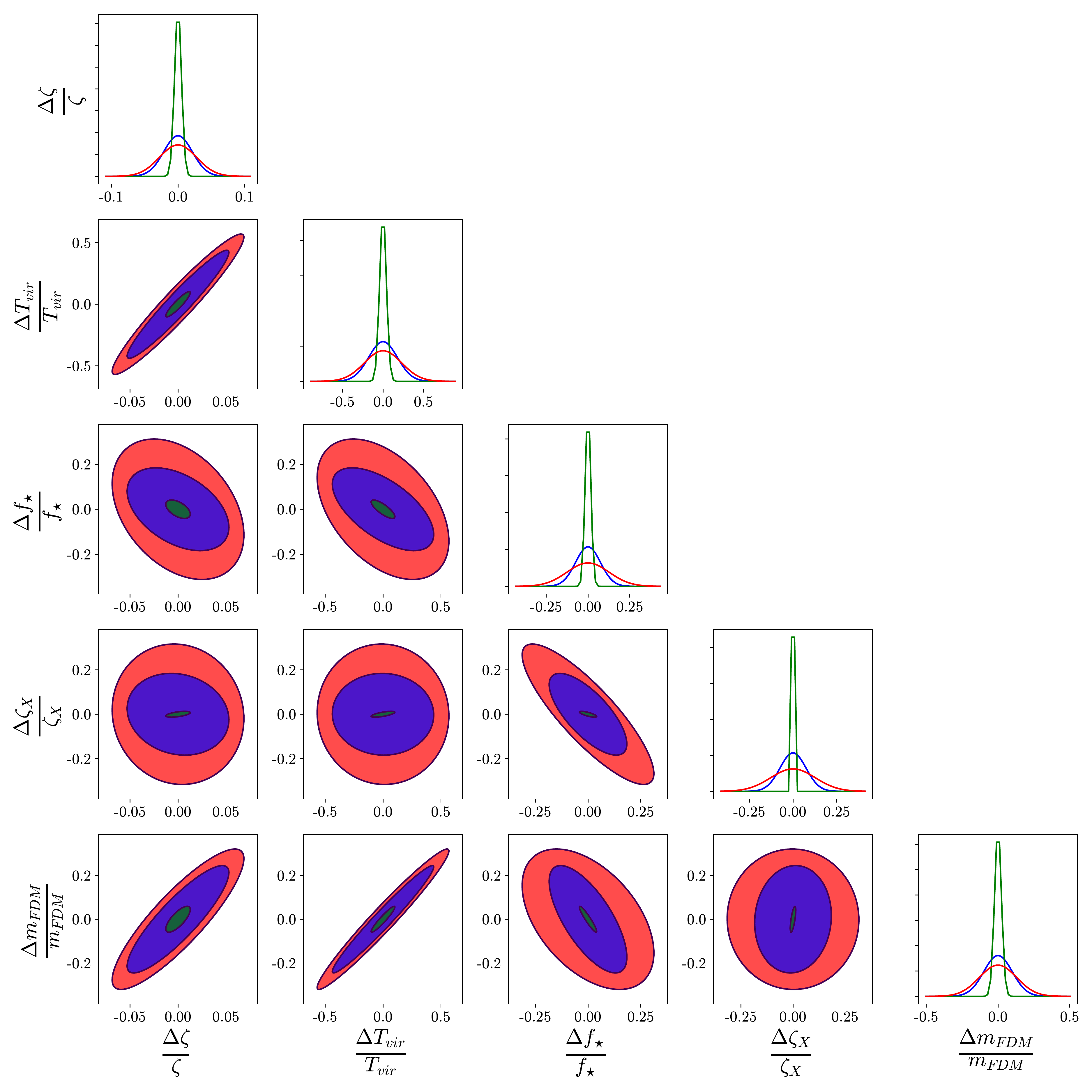}
    \caption{Parameter constraint forecasts for each of the three different foreground contamination scenarios for HERA. The optimistic, moderate, and pessimistic cases are shown in green, blue, and red, respectively. The ellipses span 2-$\sigma$ confidence intervals.} 
\label{fig:ellipses}
\end{figure}

Fig~\ref{fig:ellipses} shows the resulting parameter constraint forecasts for HERA observations in each of the pessimistic, moderate, and optimistic foreground contamination scenarios (\S \ref{sec:HERA}). These results sum over the full redshift range spanned by HERA ($z=5-27$) and over all wavenumbers. The bottom-line constraint on the FDM particle mass is given by the 1D likelihoods (marginalized over the other parameters) in the bottom-right hand panel. Encouragingly, we forecast very tight constraints on the FDM mass for all foreground treatments: HERA should determine the FDM mass to within 4.8\%, 20\%, and 26\% in the optimistic, moderate, and pessimistic scenarios, respectively at 2-$\sigma$ confidence. That is, we expect a strong {\it detection of FDM} and a tight constraint on the FDM particle mass. These numbers assume our fiducial FDM mass of $m_{FDM}=10^{-21}$ eV as the true underlying model. If we had instead assumed CDM as the fiducial model, the tight constraints shown here suggest that the upper bound on FDM mass in CDM would be significantly tighter than $10^{-21}$ eV.

As anticipated earlier, there are fairly strong parameter degeneracies between FDM mass and other astrophysical parameters. The most prominent one is with the minimum virial temperature of galaxy hosting dark matter halos. The strong positive correlation between these parameters results because increasing the FDM mass lessens the delay in the EoR from FDM, which can be counteracted by increasing $T_{\rm vir}$. The slightly less strong degeneracy seen in the $q_\zeta-q_{mFDM}$ plane is naively surprising, since we expect the errors on these parameters to be negatively correlated. This occurs, however, because $\zeta$ and $m_{FDM}$ are not the only parameters in the problem. For example, an increase in $\zeta$ can be compensated by boosting $T_{\rm vir}$ which then requires a counteracting increase in $m_{FDM}$. Indeed, if we fix all of the other nuisance parameters to their fiducial values, the degeneracy direction in the $q_\zeta-q_{mFDM}$ plane flips: in this case, these quantities show negative error correlations as naively expected. The degeneracies between the FDM particle mass and the star-formation efficiency and X-ray heating parameters are less strong. This mainly results because the error bars on HERA's power spectrum measurements during the EoR are much smaller than during the Cosmic Dawn (see \S \ref{sec:HERA}), while the star-formation efficiency and X-ray heating parameters mostly impact the Cosmic Dawn and not the EoR. 

\begin{figure}[htpb]
    \centering
     \includegraphics[width=1.0\columnwidth]{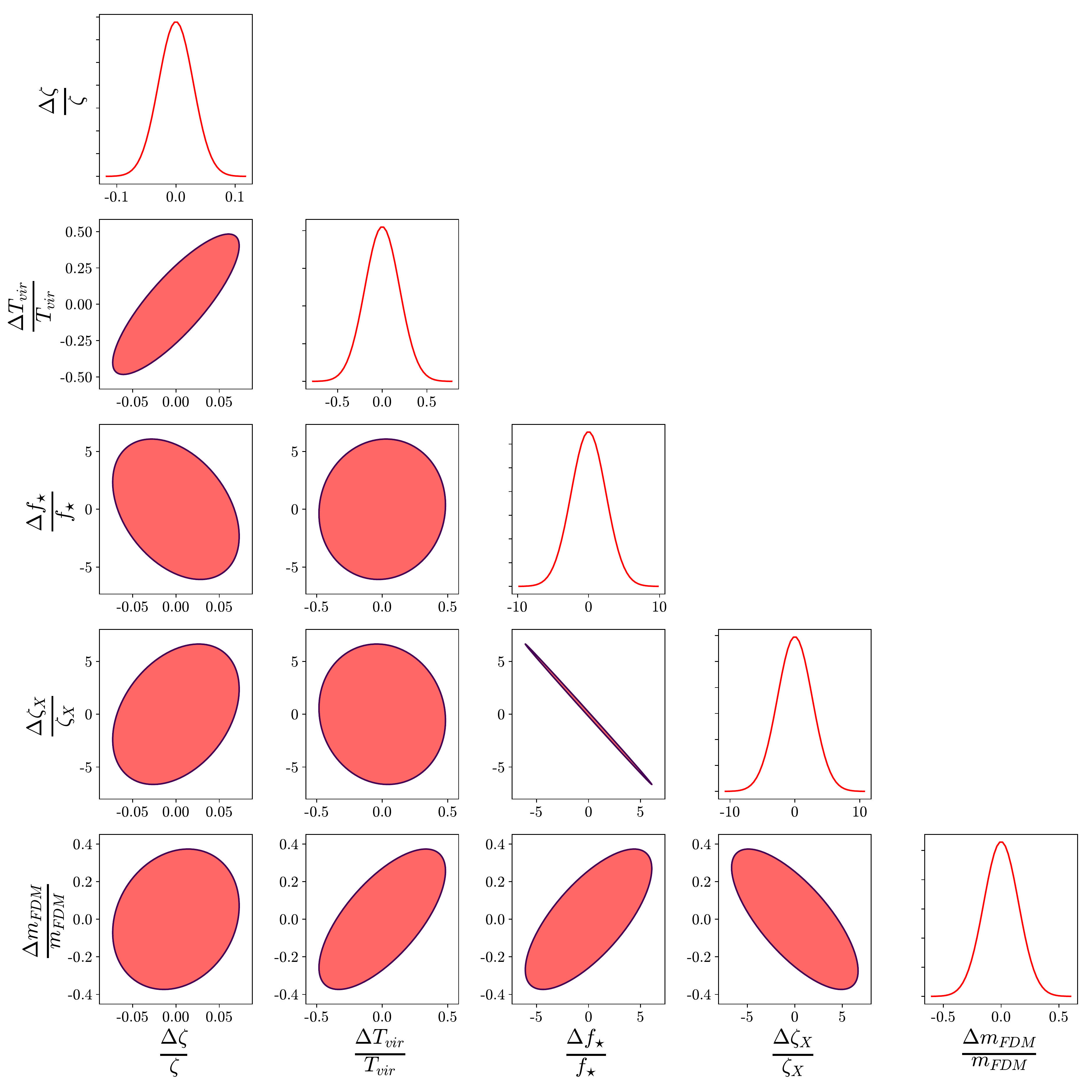}
    \caption{Constraint forecasts for HERA including only redshifts $z \leq 10$ in the Fisher matrix calculations. That is, these forecasts only include redshifts during the EoR. We adopt the moderate foreground removal scenario. The ellipses enclose $2-\sigma$ confidence regions.}
\label{fig:ellipses_EOR}
\end{figure}

\begin{figure}[htpb]
    \centering
\includegraphics[width=1.0\columnwidth]{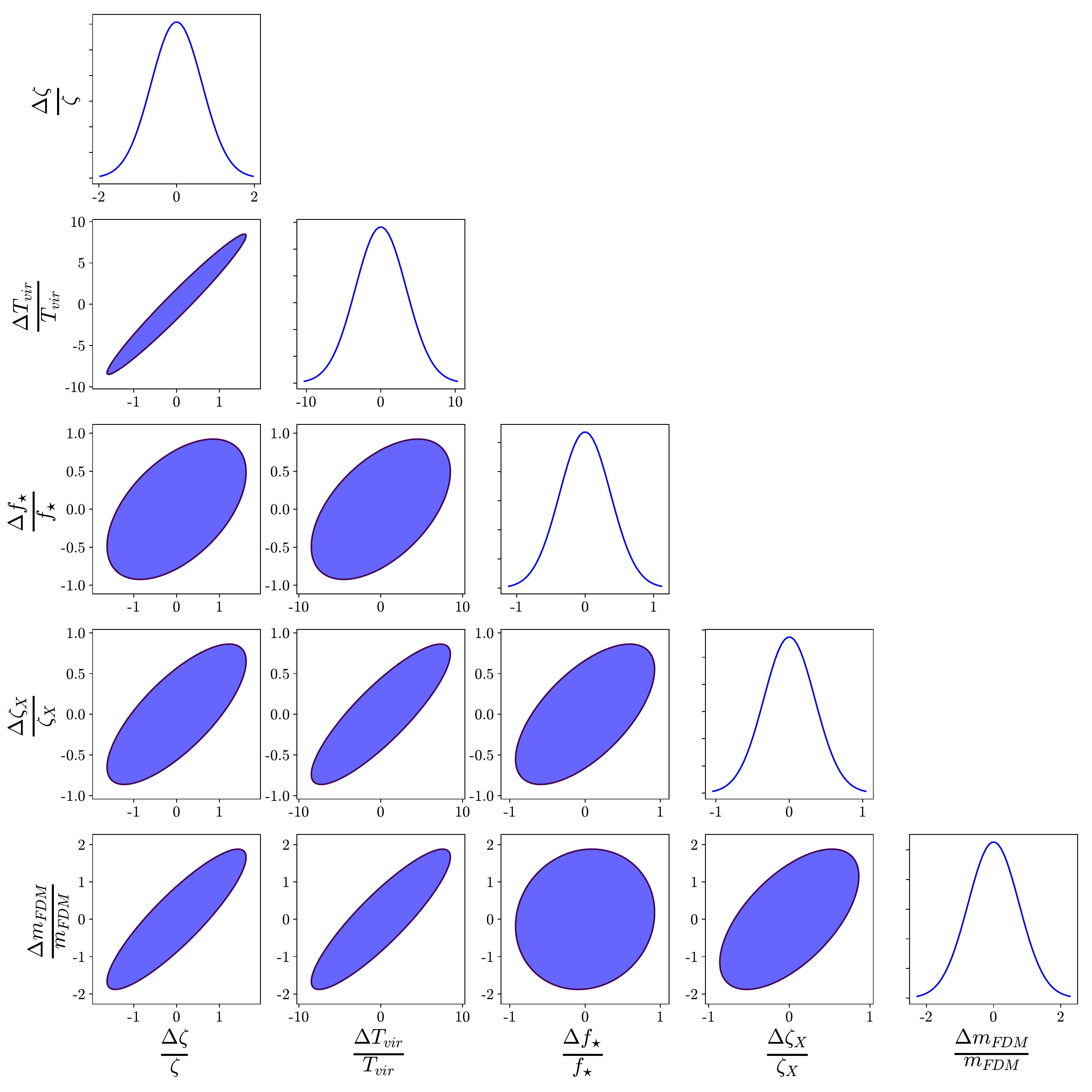}
    \caption{Constraint forecasts for HERA including only redshifts $z \geq 10$ in the Fisher matrix calculations. Here we include only Cosmic Dawn redshifts. We adopt the moderate foreground removal scenario. The ellipses enclose $2-\sigma$ confidence regions.}
\label{fig:ellipses_CD}
\end{figure}

Figs~\ref{fig:ellipses_EOR} and \ref{fig:ellipses_CD} further illustrate this by showing, respectively, the parameter constraints from the EoR and Cosmic Dawn alone. These results are shown for the moderate foreground case. The EoR calculations consider $z \leq 10$ while the Cosmic Dawn ones adopt $z \geq 10$. The FDM mass constraints from the EoR alone are a factor of $\sim 5$ tighter than those from the Cosmic Dawn alone. Although the statistical precision of the Cosmic Dawn constraints are formally much weaker, it is still appealing to constrain FDM mass with HERA power spectrum measurements in this era.
For one, somewhat different physics is involved in this period and so it provides a potential cross-check on the EoR constraints. Second, it probes the earliest stages of star and galaxy formation where FDM has an especially strong impact.
Finally, although we do not consider this combination here, one can potentially combine the power spectrum constraints with global 21 cm measurements: current global 21 cm experiments already have the sensitivity to detect the
Cosmic Dawn if systematic concerns can be mitigated (e.g. \citealt{Bowman:2018yin,Lidz:2018fqo,Schneider:2018xba,Nebrin:2018vqt,Munoz20}).

Here we can also briefly discuss how our results compare with those in the related work of \citet{Munoz20}. Both studies find that future HERA measurements should provide cutting-edge constraints on FDM. The most important difference between our study and this earlier work is that \cite{Munoz20} adopt a rather different set of fiducial astrophysical parameters. Specifically, in that study, they allow efficient star formation in molecular cooling halos with masses of order $\sim 10^6-10^7 M_\odot$, although they also include a model for dissociating Lyman-Werner band feedback which partly regulates star formation in these halos.  If star formation is indeed efficient in these small halos, the relative streaming velocity between dark matter and baryons is an important effect \citep{Tseliakhovich2010}: this leads to spatial variations in the halo-collapse fraction which in turn enhances the Cosmic Dawn era 21 cm fluctuation signal (see also e.g. \citealt{Fialkov2013}). In our model, on the other hand, we assume that star-formation is inefficient in these small mass halos, and that Ly-a coupling, X-ray heating, and reionization are accomplished entirely by stars forming in higher mass atomic cooling halos. In this case, the relative streaming velocity effect is a fairly minor one \citep{Tseliakhovich2010} and neglected here.

Their scenario hence leads to a stronger Cosmic Dawn 21 cm fluctuation signal, and so they arrive at more optimistic conclusions regarding the detectability of this era with HERA. In fact, their analysis considers only the Cosmic Dawn era signal, and not the EoR. In our model, we have seen that the constraints on FDM mass from the EoR are much stronger than those from the Cosmic Dawn (see Figures~\ref{fig:ellipses_EOR} and \ref{fig:ellipses_CD}). Their scenario requires strong evolution in the star formation efficiency towards high redshift and low halo mass (e.g. \citealt{Mirocha:2018cih}). Moreover, this case may be uncomfortable with the low electron scattering optical depths -- which bound the star formation efficiency in such halos (e.g. \citealt{Visbal:2015rpa,Miranda:2016trf}) -- suggested by Planck 2018 measurements \citep{Aghanim:2018eyx}. Hopefully, upcoming 21 cm measurements will help determine empirically which fiducial model here is more reliable. In any case, these upcoming surveys should provide interesting FDM constraints.

Finally, it is interesting to note that allowing FDM mass as a free parameter significantly degrades the constraints on the astrophysical parameters. Specifically, our constraints on $T_{\rm vir}$, $\zeta$, $f_\star$, and $\zeta_x$ decrease if we fix the FDM particle mass to the fiducial value, rather than letting it vary freely. This decrease is at the factor of several levels for $T_{\rm vir}$ and $\zeta$.
Indeed, our error forecasts on these parameter are larger than in previous work (e.g. \citealt{Ewall-Wice:2015uul}). We find very similar results to this earlier study, however, if we instead fix the FDM mass. 

\section{Conclusions}\label{sec:conclusions}

We modeled the impact of FDM on the 21 cm power spectrum and forecasted the expected constraints on FDM mass from upcoming HERA measurements. The suppression in the abundance of small mass halos leads to a delay in the Cosmic Dawn and the EoR and strongly impacts the power spectrum of 21 cm fluctuations, even for FDM models with $m_{FDM} \sim 10^{-21}$ eV that remain challenging to constrain by other means. In addition, FDM modifies the spatial structure of the 21 cm signal at a given stage of the EoR and Cosmic Dawn. This occurs because of the small mass halo suppression in FDM; the ionizing sources are hence more highly-biased tracers of the matter power spectrum in FDM than CDM.

We further characterized degeneracies between the effects of FDM and uncertain astrophysical parameters. The most important one is with the minimum virial temperature of galaxy hosting halos. In our fiducial model ($T_{\rm vir} = 2 \times 10^4$ K and $m_{FDM}=10^{-21}$ eV), however, the FDM suppression mass is larger than the minimum mass of galaxy hosting halos and so sharp constraints on FDM are still expected. In the future, measurements of the UV luminosity function with e.g. the James Webb Space Telescope may reveal a turn-over or flattening at low luminosities. The precise shape and redshift dependence of the faint end of the luminosity function may then help in separating out the effects of the minimum virial temperature and the FDM mass, especially when the UV luminosity function measurements are combined with redshifted 21 cm observations. 

Assuming an FDM model with $m_{FDM}=10^{-21}$ eV, we forecast a strong detection in upcoming HERA 21 cm observations and a tight 20\% determination of the FDM particle mass (at 2-$\sigma$ confidence). On the other hand, if CDM is the true model, we expect to strongly rule out a case with $m_{FDM} = 10^{-21}$ eV. These constraints depend on the ability of future 21 cm surveys to mitigate challenging foreground contamination systematics, but strong limits appear feasible even in the pessimistic foreground contamination scenario of \cite{Pober:2013jna}. Furthermore, we have only considered the power spectrum in this study, but more information about the 21~cm field should be contained in higher order statistics \citep[e.g., the bispectrum; ][]{Majumdar:2018}, potentially allowing even tighter constraints.

It is also interesting to note that our study has implications for warm dark matter (WDM) particle candidates. Although the transfer function in WDM has a different shape than in FDM, we can roughly translate our FDM constraints into WDM forecasts by finding the WDM particle mass that matches the suppression mass of Eq.~\ref{eq:mhalf} in FDM (see e.g. \citealt{Hui:2016ltb,Lidz:2018fqo}). In the case of thermal relic WDM, this translation gives $m_{\rm WDM} = 2.6\,{\rm keV} \left[m_{\rm FDM}/(10^{-21} {\rm eV})\right]^{0.4}$ (e.g. \citealt{Lidz:2018fqo}). Thus our fiducial FDM model roughly matches the suppression in a thermal relic WDM model with a mass of 2.6 eV. If the true model is WDM with this mass, HERA should deliver an fractional error on the WDM mass of $\sigma_{\rm m_{WDM}}/{\rm m_{WDM}} \sim 0.4\, \sigma_{\rm m_{FDM}}/{\rm m_{FDM}}$. In the moderate foreground removal scenario, for instance, this implies an $8\%$ constraint on the WDM particle mass.

Although there are uncertainties in modeling early star and galaxy formation and the resulting 21 cm signal, the redshifted 21 cm signal provides a uniquely powerful constraint on the timing of some of the earliest phases of structure formation and this gives an appealing handle on FDM models. It would be hard to reconcile FDM, or any other model in which the initial density power spectrum is suppressed on small scales, with an early start to the Cosmic Dawn and the EoR, as might be revealed via upcoming HERA measurements. The HERA observations can be combined with independent methods to convincingly rule-out or detect FDM, including: analyses of the Lyman-alpha forest \citep{Irsic:2017yje}, post-reionization 21 cm intensity mapping surveys \citep{Bauer:2020zsj}, measurements of UV luminosity functions \citep{Bozek:2014uqa}, sub-structure lensing data \citep{Dalal:2001fq}, studies of ultra-faint dwarf galaxies \citep{Marsh:2018zyw}, the potential imprint of FDM on tidal streams in our galaxy \citep{Dalal:2020mjw}, and using the black hole super-radiance phenomenon \citep{Davoudiasl19}.

\section*{Acknowledgements}
AL acknowledges support, in part, through NASA ATP grant 80NSSC20K0497. We thank the anonymous referee for helpful comments on the manuscript.

\bibliography{references}

\end{document}